\documentclass[a4paper,pre,twocolumn,superscriptaddress,showpacs]{revtex4} 

\usepackage{amsmath}
\usepackage{amssymb}
\usepackage{graphicx}
\usepackage{subfigure}
\begin{document}
\title{Moving Finite Size Particles in a Flow: \\A Physical Example for Pitchfork Bifurcations of Tori}
\author{Jens C. Zahnow}
\affiliation{Institute of Physics, University of Oldenburg, 26129 Oldenburg, Germany}
\affiliation{Theoretical Physics/Complex Systems, ICBM, University of Oldenburg, 26129 Oldenburg, Germany}
\author{Ulrike Feudel}
\affiliation{Theoretical Physics/Complex Systems, ICBM, University of Oldenburg, 26129 Oldenburg, Germany}

\begin{abstract}
The motion of small, spherical particles of finite size in fluid flows at low Reynolds numbers is described by the strongly nonlinear Maxey-Riley equations. Due to the Stokes drag the particle motion is dissipative, giving rise to the possibility of attractors in phase space. We investigate the case of an infinite, cellular flow field with time-periodic forcing. The dynamics of this system are studied in a part of the parameter space. We focus particularly on the size of the particles whose variations are most important in active, physical processes, for example for aggregation and fragmentation of particles. Depending on their size the particles will settle on different attractors in phase space in the long term limit, corresponding to periodic, quasiperiodic or chaotic motion. One of the invariant sets that can be observed in a large part of this parameter region is a quasiperiodic motion in form of a torus. We identify some of the bifurcations that these tori undergo, as particle size and mass ratio relative to the fluid are varied. This way we provide a physical example for sub- and supercritical pitchfork bifurcations of tori.
 \end{abstract}
\maketitle
\numberwithin{equation}{section}
\section{Introduction}
For a long time there have been intensive studies of the chaotic advection of point particles in flows \cite{Aref1984}-\nocite{Ottino}\cite{Tel2005}. Newer studies focus on particles of a finite size, instead of point particles without mass. Based on the early work of Basset \cite{Basset} and later Boussinesq \cite{Boussinesq} and Oseen \cite{Oseen}, equations of motion for spherical particles of a finite size in a flow at low Reynolds numbers were introduced by Maxey and Riley \cite{Maxey1983} (Maxey-Riley equations). These equations have been studied in varying context. In many cases the finite-size effects lead to completely different results compared to what is observed for ideal tracers \cite{Maxey1987}-\nocite{Benczik2002,Wang1992,Lopez2002,Benczik2003}\cite{Do2004}. In particular the results depend on the characteristics of the particles, i.e whether they are lighter or heavier than the fluid. 

One of the most important applications of these studies are problems where in addition to the dynamics of the particles in the flow active processes of the particles are included. Such active processes concern the dynamics of the particles themselves, i.e. they can change their physical and chemical properties due to various kinds of interaction. Examples are the study of biological processes (growth of plankton populations), chemical processes (chemical reactions in chaotic or turbulent flows), and physical processes (aggregation and fragmentation of marine aggregates, e.g. sediment particles in the ocean, or cloud formation) \cite{Jackson1990}-\nocite{Ruiz1997,Melis1999,Thomas1999,Flesch1999}\cite{Binbing2003}. Using the Maxey-Riley equations, the influence of the finite size of particles on such active processes as autocatalytic reactions and coalescence of particles has recently been studied in \cite{Nishikawa2001}, \cite{Nishikawa2002}. It has been shown that for particular parameters of the flow chaotic attractors can occur, leading to filamental structures of the particle distribution.
 
In this paper the Maxey-Riley equations are studied in the case of an infinite, cellular, time dependent flow field.  Due to the Strokes drag the particle motion is dissipative, giving rise to the possibility of attractors in phase space. The observed system behavior is highly complex and depends strongly on the parameter values. Different forms of periodic, quasiperiodic and chaotic behavior appear in the system, with both single attractors and parameter regions possessing multi-stability. Some particular attractors and parameter regions have already been discussed by e.g. Maxey \cite{Maxey1987}, Yu et al. \cite{Yu1990} and Nishikawa et al. \cite{Nishikawa2001}. Others have not been studied yet. Because of the complexity and the number of parameters that can be varied (a total of 6 parameters) only a part of the parameter space, namely the variation of two parameters, is studied in this paper. These two parameters are chosen for their relevance in the context of application in active, physical processes. In particular we are interested in the type of long-term behavior, which occurs for different sizes and mass of the particles. Since active processes like aggregation and fragmentation change the size of the particles, it is interesting to study the impact of variations in the size on the dynamics of the particles.

For both particles, heavier and lighter than the fluid, quasiperiodic solutions in the form of a torus can be observed in a large parameter region. In this paper we are studying some of the changes that these tori undergo, as parameters are varied. The main focus is on pitchfork bifurcations of tori that appear in this context. 

While there is a large body of work devoted to the study of bifurcations of fixed points and periodic orbits in various fields of science, there are only a few examples from application where bifurcations of tori occur. Some general results are known from the mathematical literature concerning the bifurcations of tori \cite{UnfoldingsBifurcationsofTori} and particularly the break-up of tori and the transition to chaos \cite{Newhouse1978},\cite{Afraimovich1983}. These theoretical findings are accompanied by numerical studies of paradigmatic systems \cite{Ostlund1983}-\nocite{Anishchenko1987}\cite{Baesens1991}. Examples from physics are mainly focused on the possibility of the emergence of tori with three incommensurate frequencies $T^{3}$ \cite{Held1986}-\nocite{Schneider1993,Giberti1993,Anishchenko1993}\cite{Feudel1996} and the break-up of tori with two incommensurate frequencies $T^{2}$ \cite{Franceschini1992}-\nocite{Belogortsev1993}\cite{Feudel1995}. The more basic bifurcation of a torus studied here has to our knowledge not yet been observed in a physical system and is only known from a mathematical point of view (e.g. Sun \cite{Sun2003}). For particles moving in a fluid flow two kinds of pitchfork bifurcations of a torus can be obtained in different parameter intervals. First we find a subcritical pitchfork bifurcation, where an unstable torus becomes stable and gives rise to the emergence of two new unstable tori. Second a supercritical pitchfork bifurcation is observed, where a stable torus becomes unstable and two new stable tori are created.  While the first bifurcation happens only if the particles are lighter than the fluid, the second one occurs for particles heavier than the fluid.  In contrast to the usual transition from sub- to supercritical which appears on the same bifurcation line, we find no continuous connection between the two bifurcations. 

The paper is organized as follows. In Section \ref{Equations} the equations of motions are presented. Based on the Maxey-Riley equations the complete equations of motion for small spherical particles in a time-dependent flow field are recalled and some basic properties of the system are described, following the approach of \cite{Nishikawa2002}.  Section \ref{General} gives a general overview of the system behavior depending on the size of the particles and of the part of the parameter space that is of relevance to the pitchfork bifurcation presented here. In Section \ref{SupercriticalP} the results from numerical simulations are described that show a supercritical pitchfork bifurcation of a torus. Section \ref{SubcriticalP} shows the numerical results for a subcritical pitchfork bifurcation of a torus. Section \ref{Conclusion} contains a brief summary.
  
\section{\label{Equations}System of Equations}
Our investigation is based on the motion of small rigid spherical particles in a two dimensional, time periodic flow field under the influence of gravity. The flow field consists of a regular pattern of vortices, or roll cells, that is infinitely  extended. This flow field was first introduced by Chandrasekhar \cite{Chandrasekhar} as a solution to the Benard problem, but has also been used in different context since then (e.g. \cite{WigginsGlobalChaos}, \cite{Maxey1987}). \\
The undisturbed flow field is denoted by $\vec{u}(\vec{x},t)$, where $\vec{x}=(x_{1},x_{2})$ is a position in the flow field. The flow is incompressible, with constant density $\rho$, pressure $p$ and dynamical viscosity $\eta$ of the fluid. Since we restrict ourselves to a two-dimensional flow it can be represented by a stream function
\begin{equation}\label{Psi}
\Psi(\vec{x}, t) = [1+k \sin(\omega t)]\frac{U_{0} L}{\pi} \sin(\pi x_{1}/L) \cdot\sin(\pi x_{2}/L)~.
\end{equation}
$U_{0}$ is the maximum velocity of the flow for a given time $t$, $k$ is the amplitude of the periodic forcing and $\omega$ is the frequency. The size of a vortex is $L$.

The velocity field $\vec{u}(\vec{x},t)=(u_{1}(\vec{x},t),u_{2}(\vec{x},t))$ of the fluid at the position $\vec{x}$ is derived from the stream function in Eq. \eqref{Psi} by 
\begin{equation}
u_{i} = [\nabla \times \vec{\Psi}]_{i}~,\qquad i=1,2
\end{equation}
with $\vec{\Psi} = [0, 0, \Psi]$.\\
 
The equations of motion for a particle in the flow described in Eq. \eqref{Psi} were derived by Yu et al. \cite{Yu1990}. A brief summary, following the description of Nishikawa et al. \cite{Nishikawa2002}, is shown here for completeness.

The basic equations of motion for the dynamics of a small rigid spherical particle of mass $m_{p}$ and radius $a$ in an incompressible flow for low Reynolds numbers are the Maxey-Riley equations \cite{Maxey1983}
\begin{eqnarray}\label{MaxeyRiley}
m_{p} \frac{d\vec{V}}{dt} &= &(m_{p}-m_{F})\vec{g} + m_{F}\frac{D\vec{u}}{Dt}\left.\right|_{\vec{X}(t)} \nonumber \\ 
& & - \frac{1}{2}m_{F}\frac{d}{dt}(\vec{V}-\vec{u}(\vec{X}(t),t) \nonumber \\ 
& & -\frac{1}{10}a^{2}\nabla^{2}\vec{u}(\vec{X}(t),t)) \nonumber \\ 
& &-6\pi a\eta\vec{Y}(t) \nonumber \\ 
& &-6\pi a^{2}\eta\int^{t}_{0}d\tau \frac{1}{\sqrt{\pi\nu(t-\tau)}}\frac{d}{d\tau}\vec{Y}(\tau)
\end{eqnarray}
The position of the particle is denoted by $\vec{X}=(X_{1},X_{2})$ and the particle velocity is $\vec{V}=(V_{1},V_{2})$ and $\vec{Y}(t)=\vec{V}(t)-\vec{u}(\vec{X}(t),t)-\frac{1}{6}a^{2}\nabla^{2}\vec{u}(\vec{X}(t),t)$. $m_{F}$ is the mass of the displaced fluid, $\nu$ is the kinematic viscosity and $\vec{g}=g\vec{e}_{x_2}$ is the gravitational acceleration, where $\vec{e}_{x_2}$ is the unit vector in $x_2$ direction.

These equations are derived for the conditions
\begin{eqnarray} 
aW_{0}/\nu &\ll &1 \label{Reynolds}\\
(a^{2}/\nu)(U_{0}/L) &\ll &1 \\
a/L &\ll & 1 \label{restriction3},
\end{eqnarray}
where $W_{0}$ is a representative velocity scale for ($\vec{V}-\vec{u}$). 

It is important to distinguish between 
\begin{equation}\label{AblD}
\frac{D\vec{u}}{Dt}=\frac{\partial\vec{u}}{\partial t}+(\vec{u}\cdot\nabla)\vec{u}~,
\end{equation}
the Lagrangian (substantial) derivative, taken along the trajectory of a fluid element and 
\begin{equation}\label{Abld}
\frac{d\vec{u}}{dt}=\frac{\partial\vec{u}}{\partial t}+(\vec{V}\cdot\nabla)\vec{u}~,
\end{equation}
the derivative taken along the trajectory of the particle.

The term
\begin{equation*}
m_{F}\frac{D\vec{u}}{Dt}|_{\vec{X}(t)}
\end{equation*}
is the force from the undisturbed fluid acting on the particle at position $\vec{X(t)}$.

The buoyancy force is given by
\begin{equation*}
(m_{p}-m_{F})\vec{g}
\end{equation*}

The term
\begin{equation*}
- \frac{1}{2}m_{F}\frac{d}{dt}(\vec{V}-\vec{u}(\vec{X}(t),t) - \frac{1}{10}a^{2}\nabla^{2}\vec{u}(\vec{X}(t),t))
\end{equation*}
represents the added mass effect. This term expresses the fact that an inertial particle brings a certain amount of fluid into motion too. 
 
The Stokes drag force $-6\pi a\eta\vec{Y}(t)$ is proportional to the difference between the particle velocity $\vec{V}(t)$  and the fluid velocity $\vec{u}(\vec{x},t)$.
\nocite{Feudel1992,ODE1}
The term
\begin{equation*}
-6\pi a^{2}\eta\int^{t}_{0}d\tau \frac{1}{\sqrt{\pi\nu(t-\tau)}}\frac{d}{d\tau}\vec{Y}(\tau)
\end{equation*}
is the Boussinesq-Basset history term, representing the effects of the diffusion of vorticity around the particle. This term will be neglected in the following. Manton \cite{Manton1974} showed that if fluid inertia effects are included, the Basset history term is of less significance. 

The terms with $a^{2}\nabla^{2}\vec{u}$ are the Faxen corrections for the nonuniform flow.
For the flow field $\vec{u}$ described by Eq. \eqref{Psi}, the Faxen corrections are
\begin{equation}\label{Faxen}
a^{2}\nabla^{2}\vec{u}=-\frac{2a^{2}\vec{u}}{L^{2}}~.
\end{equation}
In this case the Faxen corrections only decrease $\vec{u}$ by a small amount, which is proportional to $(a/L)^{2}$. Because of \eqref{restriction3} this term is small and will not affect the qualitative behavior of the system. The Faxen corrections are therefore neglected. 

Taking all this into consideration, the equations of motion for the inertial particles are reduced to 
\begin{eqnarray*}
(m_{p}+\frac{m_{F}}{2})\frac{d\vec{V}}{dt} &= &(m_{p}-m_{F})\vec{g}+6\pi a \eta[\vec{u}(\vec{X},t)-\vec{V}] \\
& &+m_{F}(\vec{u}\cdot\nabla)\vec{u}+\frac{1}{2}m_{F}(\vec{V}\cdot\nabla)\vec{u} \\
& &+\frac{3}{2}m_{F}\frac{\partial\vec{u}}{\partial t}~.
\end{eqnarray*}
Introducing dimensionless variables
\begin{equation}\label{Substitution}
\vec{X}^{*}=\frac{\vec{X}}{L}, \quad \vec{V}^{*}=\frac{\vec{V}}{U_{0}}, \quad \vec{u}^{*}=\frac{\vec{u}}{U_{0}}, \quad t^{*}=\frac{tU_{0}}{L}~,
\end{equation}
and defining
\begin{eqnarray}\label{eq:parameter}
R &:= &\frac{m_{F}}{m_{p}+\frac{m_{F}}{2}}\nonumber \\
A &:= &\frac{R}{2}B = \frac{R}{2}\frac{6\pi a\mu L}{\frac{1}{2}m_{F}U_{0}}  \\ 
\vec{W} &:= &\left(\frac{1}{R}-\frac{3}{2}\right)Q\vec{e}_{x_2} = \left(\frac{1}{R}-\frac{3}{2}\right)\frac{m_{F}}{6\pi a\eta U_{0}}g\vec{e}_{x_2}~. \nonumber
\end{eqnarray}
leads to the dimensionless equations of motion (the asterisks are suppressed for convenience)
\begin{equation}\label{Bewegungsgl}
\frac{d\vec{V}(t)}{dt}=A[\vec{u}-\vec{V}+\vec{W}]+R\left(\vec{u}+\frac{1}{2}\vec{V}\right)\cdot\nabla\vec{u}+\frac{3}{2}R\frac{\partial\vec{u}}{\partial t}~.
\end{equation}
The parameter $A$ represents the effect of the particle inertia. R is the mass ratio parameter between the fluid and the particle. Systems with $R<\frac{2}{3}$ correspond to particles heavier than the fluid (aerosols) and $R>\frac{2}{3}$ corresponds to particles lighter than the fluid (bubbles). $R\rightarrow0$ is the aerosol limit, where $m_{p}$ tends to infinity, compared to $m_{F}$. $R\rightarrow2$ is the bubble limit, where $m_{p}$ tends to 0, compared to $m_{F}$. $R=\frac{2}{3}$ corresponds to particles which are neutrally buoyant. The parameter $\vec{W}$ is the scaled particle settling velocity for still fluid. $\vec{W}$ is positive for bubbles and negative for aerosols.

Substituting
\begin{eqnarray*}
\omega * &= &\frac{\omega L}{U_0}\\
 \Psi^{*} &= &\frac{\Psi}{U_0 L}
\end{eqnarray*}
and again suppressing the asterisks yields the dimensionless stream function
\begin{equation}\label{Psi*}
\Psi(x_{1}(t),x_{2}(t),t)=\frac{1}{\pi}[1+k\sin(\omega t)]\sin(\pi x_{1})\sin(\pi x_{2})~.
\end{equation}

The dimensionless velocity field can then be derived as
\begin{eqnarray} \label{velocityfield}
\vec{u}(\vec{x}(t),t) &= & \left(\begin{array}{c}
\frac{\partial\Psi(\vec{x}(t),t)}{\partial x_{2}} \\
-\frac{\partial\Psi(\vec{x}(t),t)}{\partial x_{1}}
\end{array}\right) \\
& = & [1+k\sin(\omega t)]\left(\begin{array}{c}
\sin(\pi x_{1})\cos(\pi x_{2}) \nonumber \\ 
-\cos(\pi x_{1})\sin(\pi x_{2})
\end{array}\right)~. 
\end{eqnarray}
Combining \eqref{velocityfield} with Eq. \eqref{Bewegungsgl} results in the full equations of motion
\begin{eqnarray}
\frac{dX_{1}}{dt} &= &V_{1} \label{Bewgl1}\\
\frac{dX_{2}}{dt} &= &V_{2} \label{Bewgl2}
\end{eqnarray}
\begin{eqnarray}
\frac{dV_{1}}{dt} &= &-AV_{1}+A(1+k\sin(\omega t))\sin(\pi X_{1})\cos(\pi X_{2}) \nonumber \\
 & &+\frac{R}{2}\pi(1+k\sin(\omega t))(V_{1}\cos(\pi X_{1})\cos(\pi X_{2})\nonumber\\
 & &-V_{2}\sin(\pi X_{1})\sin(\pi X_{2}))\nonumber \\
& &+\frac{1}{2}R\pi(1+k\sin(\omega t))^{2}\sin(\pi X_{1})\cos(\pi X_{1}) \nonumber \\
& &+\frac{3}{2}R\omega k\cos(\omega t)\sin(\pi X_{1})\cos(\pi X_{2})\label{Bewgl3} \\
\frac{dV_{2}}{dt} &= &-AV_{2}-A(1+k\sin(\omega t))\cos(\pi X_{1})\sin(\pi X_{2}) \nonumber \\
 & &+\frac{R}{2}\pi(1+k\sin(\omega t))(V_{1}\sin(\pi X_{1})\sin(\pi X_{2})\nonumber\\
 & &-V_{2}\cos(\pi X_{1})\cos(\pi X_{2})) +AW\nonumber \\
& &+\frac{1}{2}R\pi(1+k\sin(\omega t))^{2}\sin(\pi X_{2})\cos(\pi X_{2}) \nonumber \\
& &-\frac{3}{2}R\omega k\cos(\omega t)\cos(\pi X_{1})\sin(\pi X_{2})\label{Bewgl4}~.
\end{eqnarray}

The dynamics of the system Eqs. \eqref{Bewgl1} - \eqref{Bewgl4} takes place in a five dimensional phase space (two positions, two velocities and time). Because of the periodic forcing we consider a stroboscopic map $M$ of the system with the period $T$ of the forcing. This map projects the dynamics onto a four dimensional phase space.

The dynamics in this four dimensional phase space is dissipative. Calculation of the divergence of the flow defined by Eqs. \eqref{Bewgl1} - \eqref{Bewgl4} yields that an element of the phase space $\mathbb{R}^{4}$ shrinks exponentially with $e^{-2At}$. Therefore, attractors can be found in this system. These attractors occur in the whole phase space, but in the following only their projections onto the configuration space are shown.

The dimensionless stream function Eq. \eqref{Psi} has a spatial period of 2. This spatial period results in a periodic structure of the phase space, with the same period. This periodic structure of the phase space is reflected in the invariant sets. It can be easily seen that for every invariant set of Eqs. \eqref{Bewgl1} - \eqref{Bewgl4} there is an infinite number of identical invariant sets, repeated with the same spatial period 2 throughout the phase space. Furthermore, if $(X_{1}(t),X_{2}(t),V_{1}(t),V_{2}(t))$ is an invariant set of Eqs. \eqref{Bewgl1} - \eqref{Bewgl4}, an identical invariant set can be found with each of the transformations $(X_{1}\rightarrow 2-X_{1}; V_{1}\rightarrow -V_{1})$ and $(X_{1}\rightarrow 1-X_{1};X_{2}\rightarrow 1+X_{2}; V_{1}\rightarrow -V_{1})$ and $(X_{1}\rightarrow 1+X_{1};X_{2}\rightarrow 1+X_{2})$. It is therefore sufficient to consider the map $M$ restricted to the unit cell $F=[0,1]\times[0,1]\times\mathbb{R}^{2}$, with $\mathbb{R}^{2}$ being the velocity components.  Using the transformations above, every invariant set in $F$ can be extended to the whole phase space. In the figures shown in this paper, the invariant sets are already extended from the unit cell $F$ to the interval $\tilde{F}=[0,1]\times[0,2]\times\mathbb{R}^{2}$. By extending the sets to $\tilde{F}$ it is easier to see what the invariant sets look like in the whole phase space, because starting with $\tilde{F}$ the sets can be extended in $X_{1}$ direction by a simple reflection along the axes $X_{1}=0,1$ and in $X_{2}$ direction by identifying $X_{2}=0$ and $X_{2}=2$. Only the projections of this interval onto the configuration space $(X_{1},X_{2})$ are shown.

In addition to the parameters $R$, $A$ and $W$ we introduce a size class parameter $\alpha$ for the particles. The size class parameter $\alpha$ describes how the radius and mass of a particle are connected to the radius and mass of a basic particle with radius $a_{0}$ and mass $m_{0}$. We assume that all particles are spherical and have a mass and volume that are the sum of the masses and volumes of a number $\alpha$ of the basic particles. For the radius and the mass of a particle of class $\alpha$ this yields $a_{\alpha}=\sqrt[3]{\alpha}\cdot a_{0}$ and $m_{\alpha}=\alpha\cdot m_{0}$ (with $m=m_{p}$ or $m=m_{F}$).  For the equations of motion of the particles this results in a change of the parameter $A$ to $A_{\alpha}=\alpha^{-\frac{2}{3}} A$ and a change of the parameter $W$ to $W_{\alpha}=\alpha^{\frac{2}{3}} W$. 

 In the following the parameter values $B=6.4$, $Q=-1.6$, $k=2.72$ and $\omega=\pi$ are chosen to allow a comparison with results by e.g. Nishikawa et al. \cite{Nishikawa2002}. The period of the forcing is therefore $T=2$. The parameters $R$ and $\alpha$ are varied. Additionally $A$ and $\vec{W}$ change accordingly when varying $R$ and $\alpha$.

\section{\label{General}General behavior}
According to the aim of our study we are interested in the qualitative behavior of the nonlinear system described by Eqs. \eqref{Bewgl1} - \eqref{Bewgl4}. Depending on the system parameters we can distinguish different kinds of long-term dynamics, such as periodic orbits, quasiperiodic and chaotic motion. Our goal is to identify the occurring invariant sets, both stable and unstable ones, and to analyze the bifurcations leading to changes in the long-term behavior. As already mentioned in the introduction, aggregation and fragmentation change the size of the particles. Even though we do not take these processes explicitly into account and deal only with the dynamics of passive tracers we focus on the size of the particles as the most important bifurcation parameter. We find that the dynamics of the particles depends strongly on their size, i.e. different invariant sets are obtained for different sizes. Additionally we obtain an interesting bifurcation that we study in more detail. Before investigating this bifurcation we present the overall picture of qualitative behavior in some parts of the parameter space.

\begin{figure}[htb]
		\centering
		\subfigure{\label{fig:R1_bifurkation_lang}\includegraphics[width=0.47\textwidth, height=5.1cm]{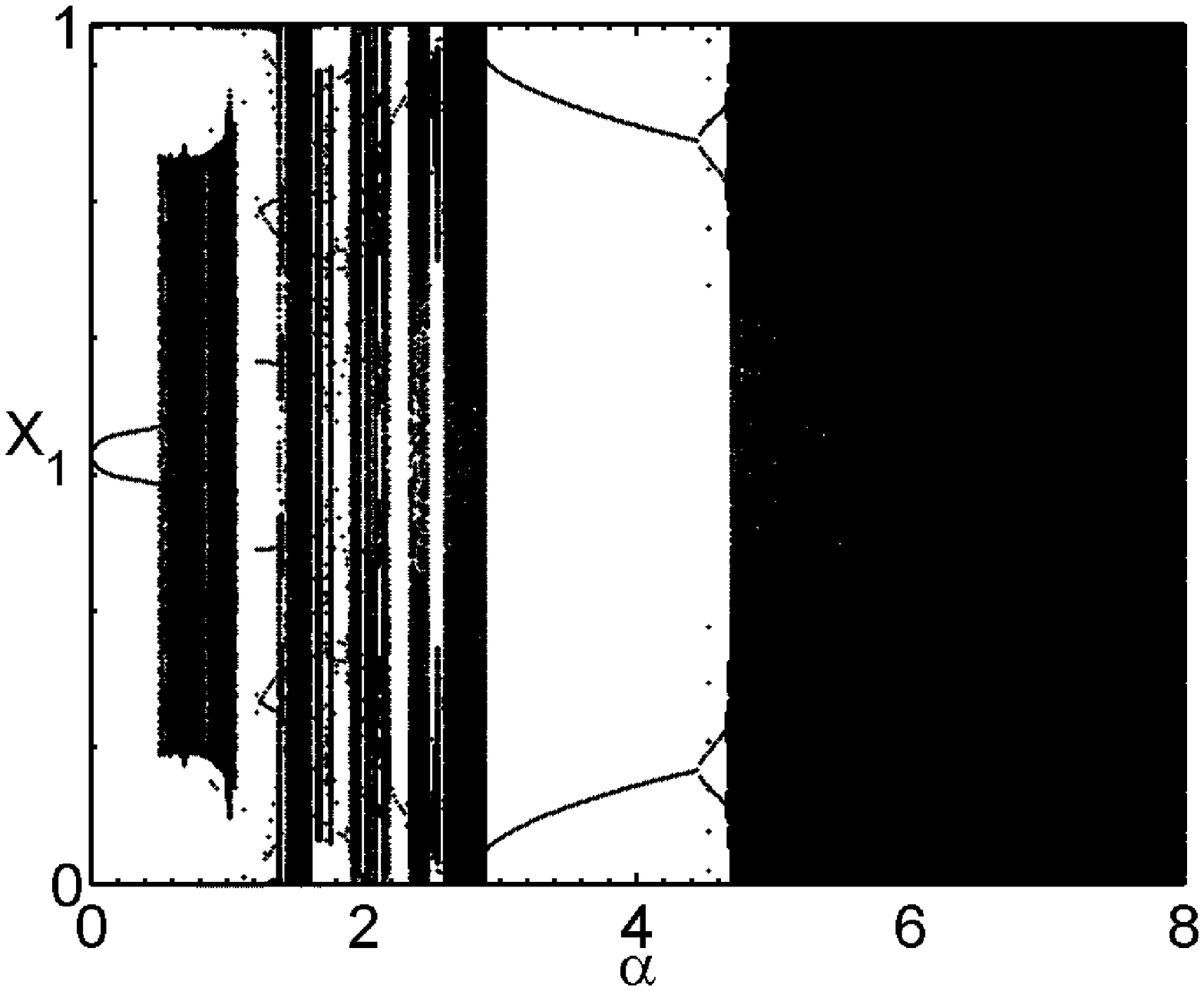}}\\
		\subfigure{\label{fig:R1_Lyap12} \includegraphics[width=0.47\textwidth, height=5.1cm]{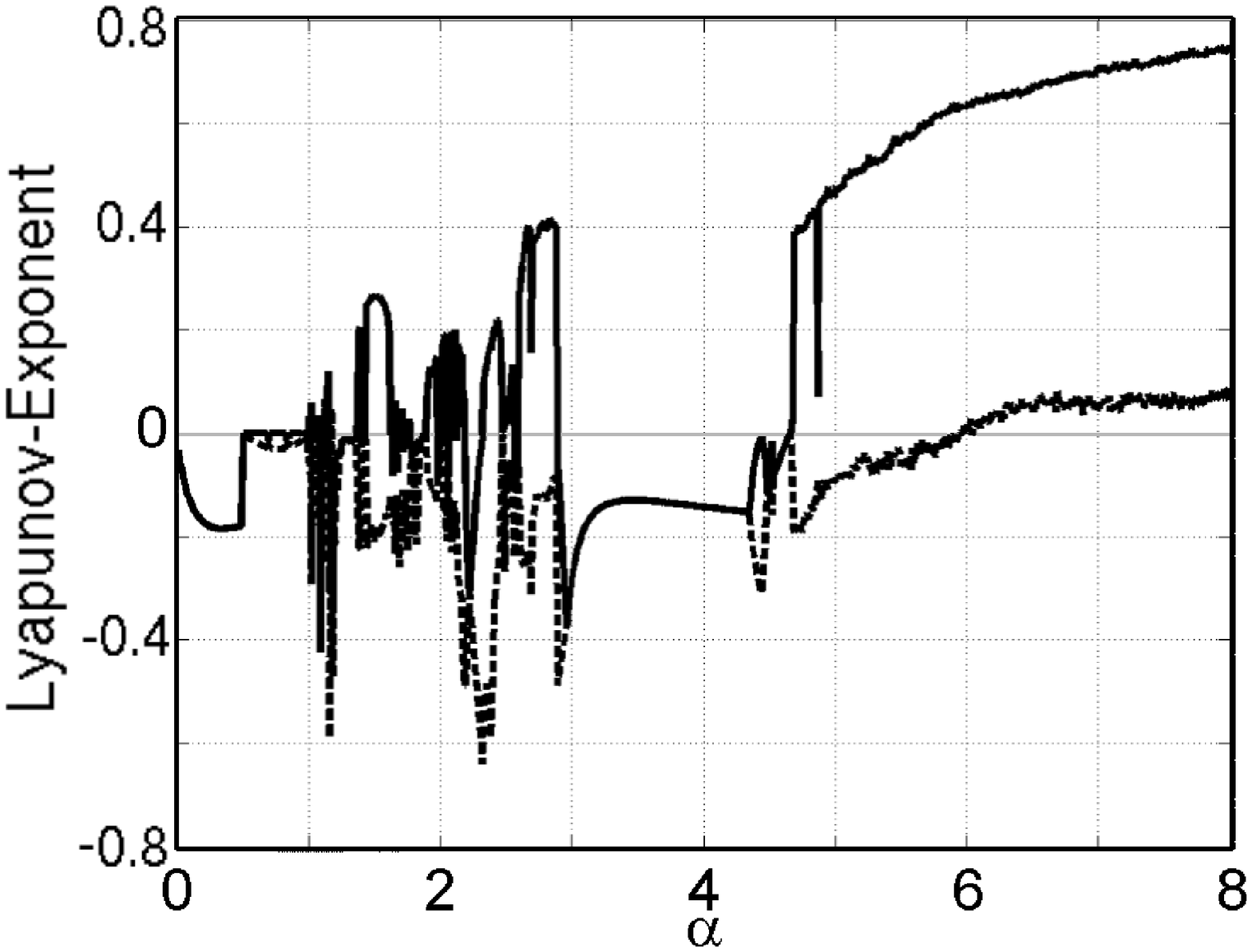}}
		\caption{\label{fig:Bif_Lyapunov_R1} Bifurcation diagram for particles lighter than the fluid (mass ratio parameter $R=1$). Upper Panel: Projection of attractors in one dimension of configuration space. Lower Panel: Numerical estimates of the two largest Lyapunov exponents $\lambda_{1}$ (solid line) $\lambda_{2}$ (dashed line).}	
\end{figure}

Since the system can not be solved analytically, numerical calculations are required to find the invariant sets corresponding to different long-term dynamics. Attractors can be easily found by examining the long-term simulations of the particle dynamics. Unstable invariant sets are harder to find due to their saddle type character. Estimates are made by starting very close to the unstable invariant sets and then examining the short-term behavior. To get a starting point close to the unstable invariant set, a refinement procedure has been applied to minimize the distance to the unstable invariant set. We follow the basic idea of the PIM-triple method which has been developed to compute chaotic saddles \cite{Nusse-Yorke-89}. We start with a set of initial conditions along a line with fixed $X_2$, $V_1$ and $V_2$ but varying $X_1$. Since the unstable invariant set and its stable manifolds separate the basins of attraction of different stable invariant sets, we integrate all initial conditions until they reach one of the attractors. Then we take the initial condition, say $I_2$, which (i) lies between two initial conditions $I_1$ and $I_3$ converging to different attractors and (ii) possesses the maximum time needed to reach the attractor (PIM-proper interior maximum for the transient time). This point $I_2$ is then taken as a starting point for the unstable invariant set. If the transient time is not long enough, so that $I_2$ is not close enough to the invariant set, then the whole procedure is repeated with a line of initial conditions connecting $I_1$ and $I_3$.

\begin{figure}[htb]
		\centering
		\subfigure{\label{fig:R06_bifurkation_lang}\includegraphics[width=0.47\textwidth, height=5.1cm]{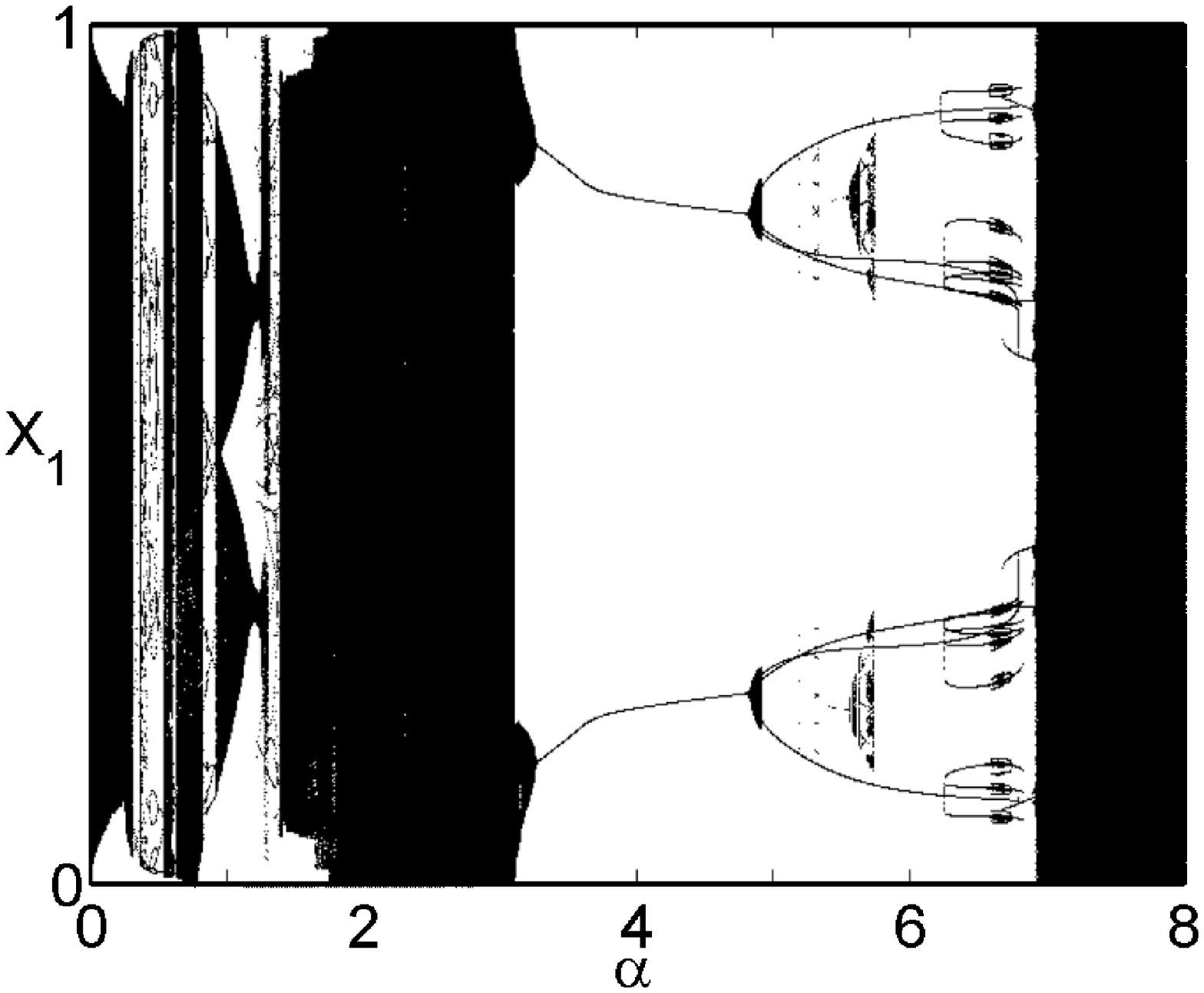}}\\
		\subfigure{\label{fig:R06_Lyap12} \includegraphics[width=0.47\textwidth, height=5.1cm]{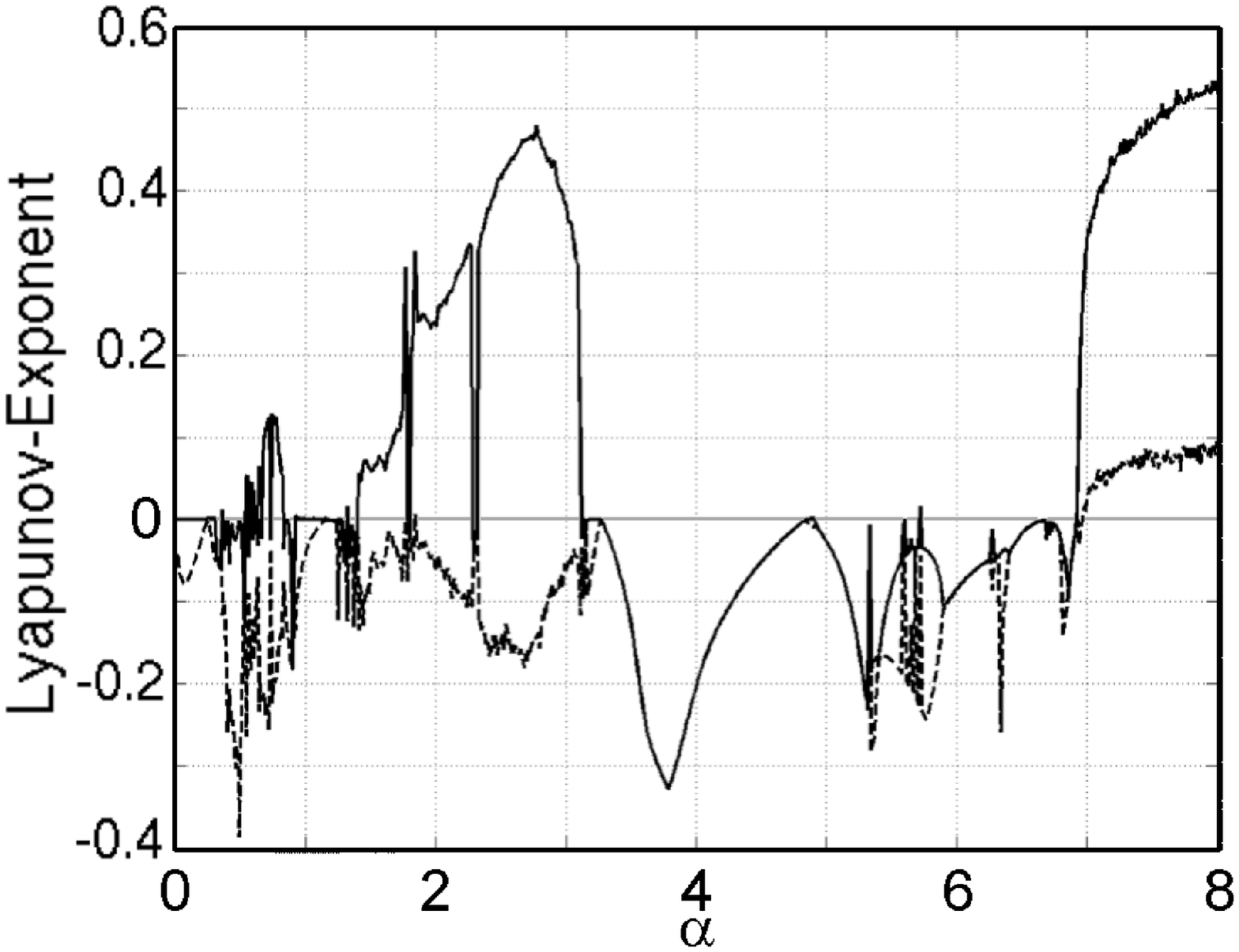}}
		\caption{\label{fig:Bif_Lyapunov_R06}  Bifurcation diagram for particles heavier than the fluid (mass ratio parameter $R=0.5$). Upper Panel: Projection of attractors in one dimension of configuration space. Lower Panel: Numerical estimates of the two largest Lyapunov exponents $\lambda_{1}$ (solid line) $\lambda_{2}$ (dashed line).}	
\end{figure}

The qualitative behavior of the system is highly dependent on the parameter values. In total there are 6 parameters involved that can be varied. Because of the number of different invariant sets that occur for variations of each of these parameters no attempt is made to give a complete description of all the invariant sets and bifurcations occurring in the whole parameter space. Thus we focus on the variation of a few parameters and keep the others fixed. 

As already mentioned the size class parameter $\alpha$ will be the main control parameter to be varied. In addition we study the behavior for particles with different mass ratios compared to the fluid, i.e. for different values of $R$. According to Eq. \eqref{eq:parameter} the variation of $R$ also implies a variation of the parameters $A$ and $\vec{W}$.

In Fig. \ref{fig:Bif_Lyapunov_R1} (upper panel) and Fig. \ref{fig:Bif_Lyapunov_R06} (upper panel) bifurcation diagrams (varying $\alpha$) for two different values of $R$ are shown. Fig. \ref{fig:Bif_Lyapunov_R1} is in the bubble region, with $R>\frac{2}{3}$, while Fig. \ref{fig:Bif_Lyapunov_R06} is in the aerosol region, with $R<\frac{2}{3}$. Both bifurcation diagrams look rather complicated. For different size classes $\alpha$ we observe different long-term dynamics, in particular periodic, quasiperiodic and chaotic motion. Varying the size classes we find various bifurcations, indicating transitions between different types of motion. Looking only at the projections of the trajectories it is not possible to distinguish between quasiperiodic and chaotic motion. Such a distinction requires the computation of the Lyapunov exponents, which are presented in Fig. \ref{fig:Bif_Lyapunov_R1} (lower panel) and \ref{fig:Bif_Lyapunov_R06} (lower panel) respectively. 
The computation of the Lyapunov exponents is based on the algorithm developed by Shimada and Nagashima \cite{Shimada-Nagashima-79} which
allows for the computation of the whole spectrum of Lyapunov exponents separately for each attractor occurring in the system. It is
important to note that the figure showing the Lyapunov exponents indicates for each parameter value always the largest occurring
Lyapunov exponents in the system, computed for a set of 30 different initial conditions. This means that even though the Lyapunov exponents are shown as a smooth curve, they can correspond to different attractors. Such a visualization has been chosen to emphasize always the most complex motion which can be found for a particular parameter value. 

Based on simulations and Lyapunov exponents we can immediately identify period doubling, e.g. at $R=1, \alpha\approx4.44$ and $R=0.5, \alpha\approx6.8$, transitions to chaos, e.g at $R=1, \alpha\approx4.68$ and $R=0.5, \alpha\approx6.92$ and torus bifurcations, e.g. at $R=1, \alpha\approx0.5$ and $R=0.5, \alpha\approx0.248$ (compare Fig \ref{fig:R06_bifurkation} (lower panel) and Fig. \ref{fig:R1_bifurkation} (lower panel)). Intermittency can be expected at e.g. $R=1, \alpha\approx2.85$. This is closely related to the emergence of periodic windows within the chaotic parameter ranges. 

\begin{figure}[htb]
		\centering
		\includegraphics[width=0.5\textwidth, height=5.3cm]{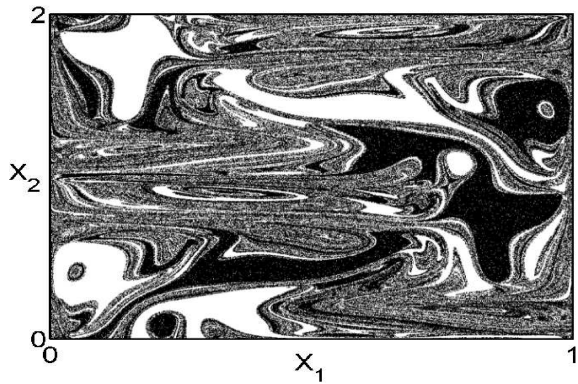}
		\caption{\label{fig:Basins_complex} Black and white dots indicate basins of attraction of two different period-2$T$ orbits for $R=1$, $\alpha=4$. The period-2$T$ orbits occur in the whole region $[0,2]\times[0,2]$, i.e. particles on these orbits move from one unit cell to the next (in $x_1$ direction) and then back. The basins of attraction display a complexly interwoven structure. To emphasize the complicated fractal structure we show in the figure only the part $[0,1] \times[0,2]$ of the configuration space, the part $[1,2]\times[0,2]$ is symmetric to the one shown.}	
\end{figure}

Within the considered parameter range we find regions where only a single attractor occurs and regions with multistability where several attractors coexist for the given set of parameter values, as i.e. seen for $\alpha \in [5.2,5.7]$ for $R=0.5$. In the latter regions it depends crucially on the initial conditions which of these stable states is realized. Each of these coexisting attractors has its own basin of attraction. For many of the considered parameter ranges these basins have a complexly interwoven structure leading to a very high sensitivity of the final state to initial conditions (see for example Fig. \ref{fig:Basins_complex}). In the following the basins of attraction shown are computed as two-dimensional cross-sections of the complete four-dimensional basins of attraction for initial conditions with $(X_1,X_2) \in [0,1] \times [0,2]$ and $(V_1,V_2)=(0,0)$.  

From the physical point of view we note that particles of different size behave differently in the same flow, i.e. their motion is on distinct attractors. As a consequence one will find the particles belonging to different size classes at different locations in the flow. We also note that bubbles and aerosols of the same size exhibit different behavior as well.

From all the occurring bifurcations we choose the torus pitchfork bifurcation for a detailed analysis, since to our knowledge it has not been discovered in this context of application so far. Therefore, we restrict ourselves to the parameter region $\alpha \in[0, 1]$ and $R\in[0.4,1.1]$. This parameter interval covers parts of the bubble region and of the aerosol region in parameter space, where tori and their bifurcations can be observed. Both a supercritical and a subcritical pitchfork bifurcation of these tori can be observed. Note that we study the stroboscopic map $M$ in which the quasiperiodic motion on a torus in the original flow appears as an invariant curve.

In the supercritical pitchfork bifurcation a stable torus looses its stability and two new stable tori are created. Such a bifurcation has been studied in the quasi-periodically forced circle map \cite{Osinga2001}. There, this bifurcation simply occurs because the quasiperiodic forcing turns the normal pitchfork bifurcation of fixed points (periodic orbits of period 1) into a pitchfork bifurcation of invariant curves in the map, corresponding to tori in flows.
\begin{figure}[htb]
		\centering
		\includegraphics[width=0.47\textwidth, height=5.1cm]{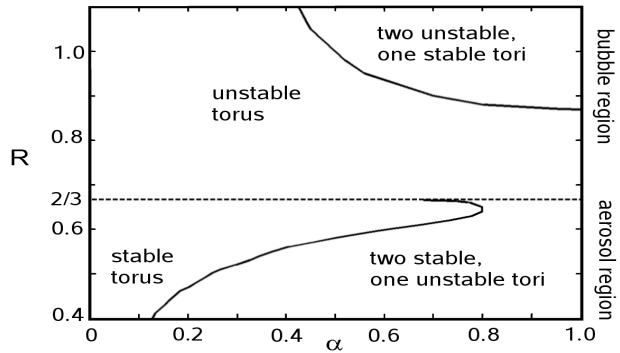}
		\caption{\label{fig:Bifurkation_reg_all} Part of the parameter space spanned by size class parameter $\alpha$ and the mass ratio parameter $R$, where the pitchfork bifurcation of a torus occurs. Solid lines indicate a pitchfork bifurcation (supercritical in the aerosol region, subcritical in the bubble region). The dashed line marks a torus bifurcation at the border between aerosol and bubble region (at $R=2/3$).}	
\end{figure}

The subcritical pitchfork bifurcation of tori follows a different scenario. An unstable torus becomes stable and two unstable tori are created which coexist with the stable torus. Though proofs of existence for both of these bifurcations are known from mathematical literature (Broer et al. \cite{UnfoldingsBifurcationsofTori}, Sun \cite{Sun2003}) they have not been analyzed in detail in the context of a physical application. 

Let us now discuss the occurrence of the pitchfork bifurcation of tori in the two dimensional parameter space spanned by $R$ and $\alpha$.
First we note that the pitchfork bifurcation of tori occurs in the bubble region as well as in the aerosol region, but there is a fundamental difference between them. In the aerosol region ($R<\frac{2}{3}$) the pitchfork bifurcation of a torus is always supercritical, giving rise to the emergence of two stable tori. The pitchfork bifurcation of a torus in the bubble region ($R>\frac{2}{3}$) is always subcritical, with two unstable tori branching off. 

There is a gap between the two bifurcation curves marked by the line $R=\frac{2}{3}$, where the particles are neutrally buoyant, so that there is no continuous transition from sub- to supercritical as known for many other bifurcations. The two stable tori that are found in the aerosol region after the pitchfork bifurcation disappear in a Neimark-Sacker bifurcation at $R=\frac{2}{3}$ when $R$ is increased (dashed line in Fig. \ref{fig:Bifurkation_reg_all}). There each of the stable tori merges with an unstable periodic orbit. At that point the tori disappear and the periodic orbits become stable. At the same point the spatial symmetry of the unstable torus is reversed (compare Fig. \ref{fig:R06_beforeafter} and \ref{fig:R1_beforeafter}).

The different cases occurring in this parameter region are described in detail in the following section. Both the stable and the unstable invariant sets are estimated numerically and their projections onto the configuration space are shown.

\section{\label{Pitchfork}Pitchfork bifurcation of tori}
\subsection{\label{SupercriticalP}Supercritical pitchfork}
\begin{figure}[htb]
		\centering
		\subfigure{\label{R05_bif_close}\includegraphics[width=0.47\textwidth, height=5.1cm]{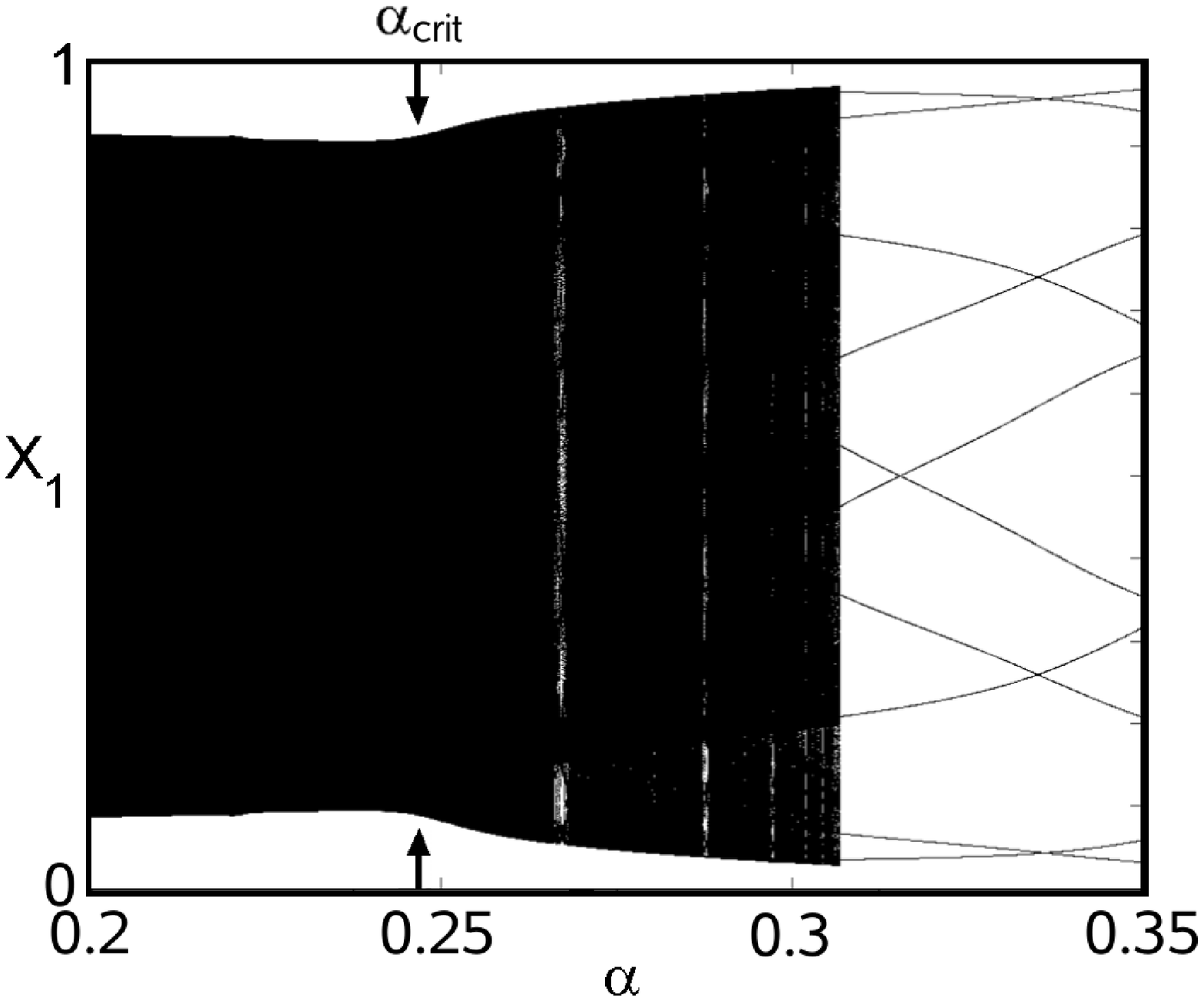}}\\
		\subfigure{\label{R05_lyap_close}\includegraphics[width=0.47\textwidth, height=5.1cm]{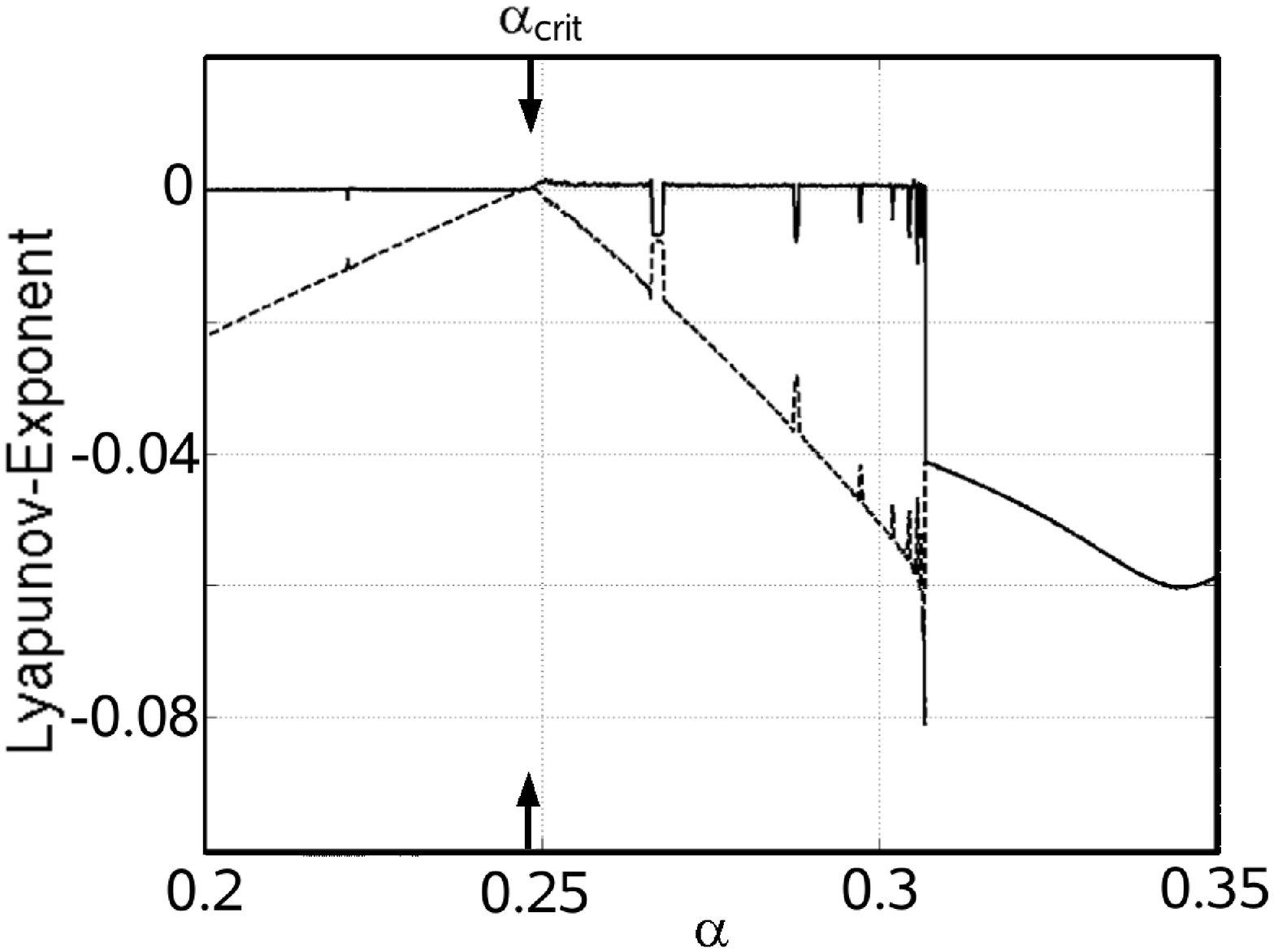}}
		\caption{\label{fig:R06_bifurkation} Bifurcation diagram and numerical estimates of the two largest Lyapunov exponents for $R=0.5$, in the area of the supercritical pitchfork bifurcation. The arrow indicates the bifurcation point at $\alpha_{crit}=0.2475$. Beyond the bifurcation point some larger periodic windows, corresponding to phase-lockings on the tori, can be easily seen. These periodic windows appear where the largest Lyapunov exponent drops suddenly below $0$, e.g at $\alpha\approx0.268$. This is also visible in the bifurcation diagram (upper panel).}	
\end{figure}
\begin{figure}[htb]
		\centering
		\subfigure{\label{fig:R06_alpha03} \includegraphics[width=0.47\textwidth, height=5.1cm]{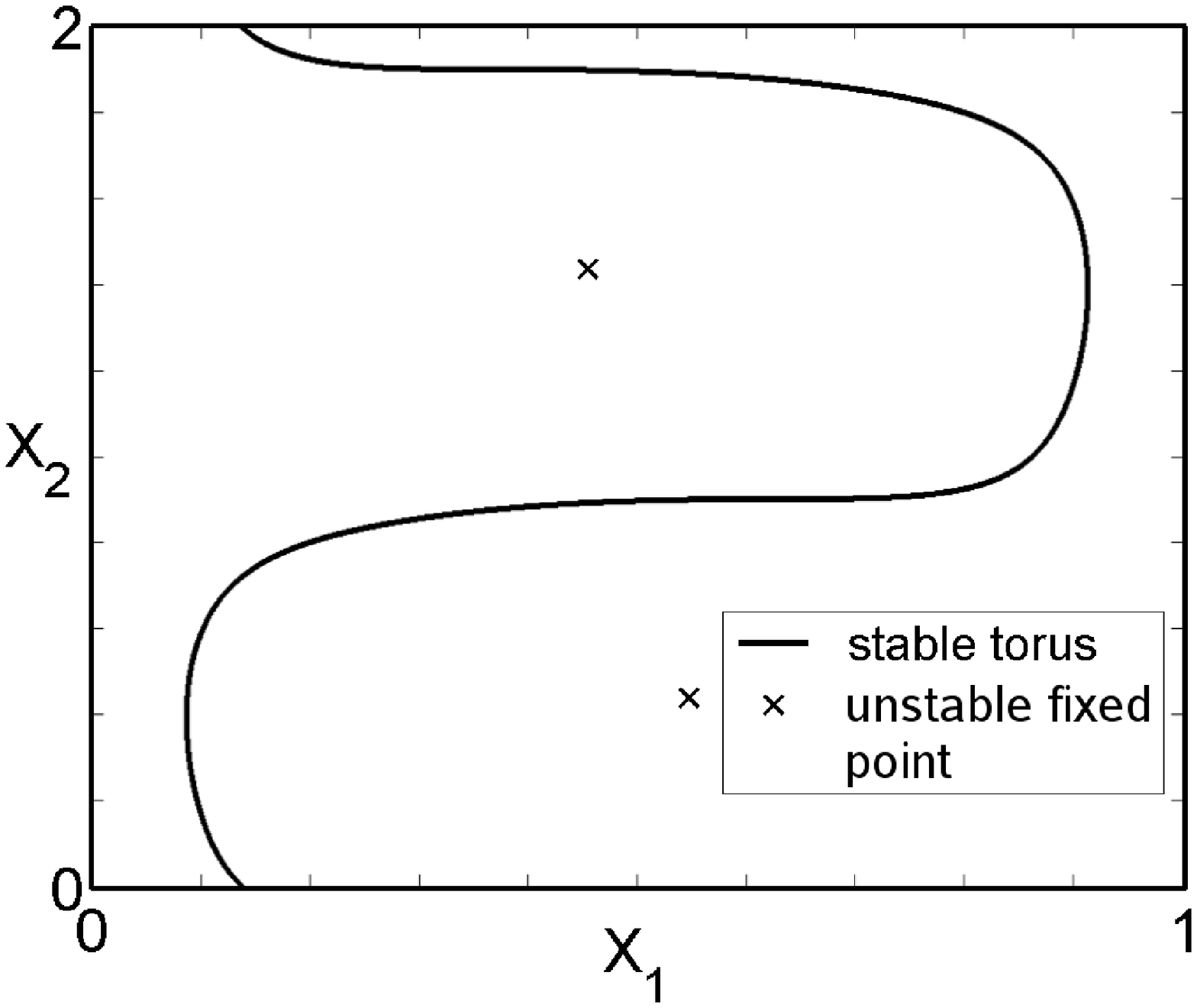}}\\
		\subfigure{\label{fig:R06_alpha0325} \includegraphics[width=0.47\textwidth, height=5.1cm]{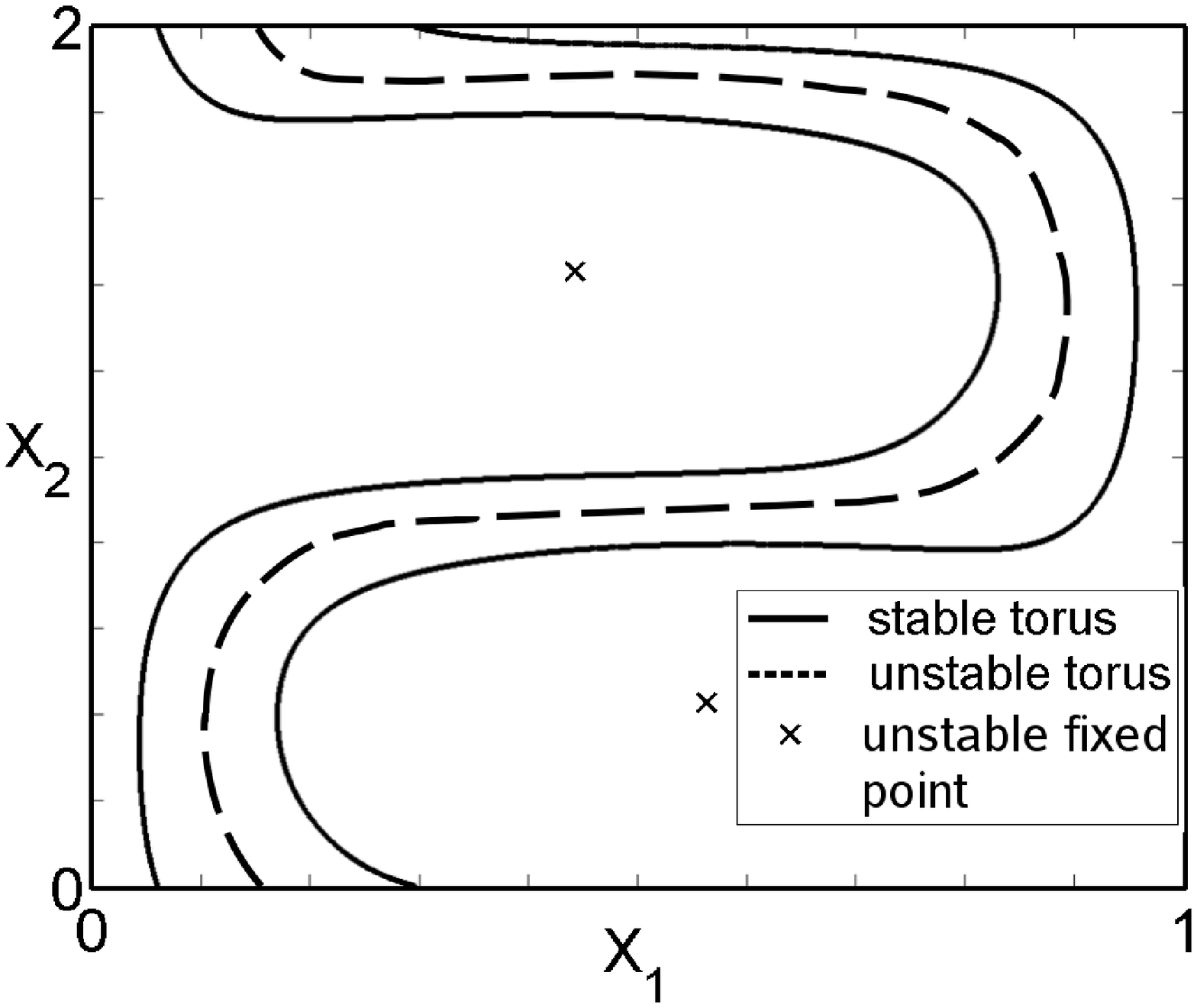}}
		\caption{\label{fig:R06_beforeafter} Supercritical pitchfork bifurcation of a torus at $R=0.5$ (solid line - stable torus, dashed line - unstable torus, cross - unstable periodic orbit). Upper panel: Before the bifurcation ($\alpha=0.2$). Lower panel: After the bifurcation ($\alpha=0.28$).}	
\end{figure}

\begin{figure}[htb]
		\centering
		\includegraphics[width=0.47\textwidth, height=5.1cm]{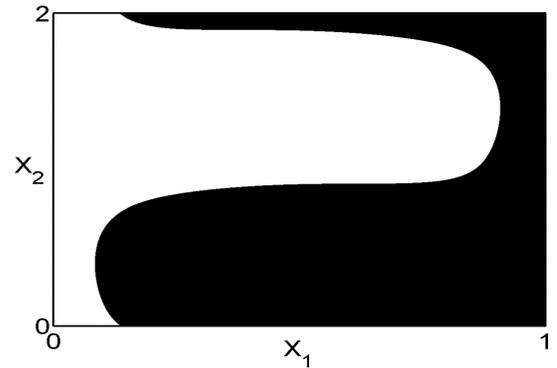}
		\caption{\label{fig:R06_alpha0325_einz} Basins of attraction for the two stable tori for $R=0.5$, $\alpha=0.325$. The basins of attraction are separated by an unstable torus.}	
\end{figure}

In this section the aerosol region of the part of the parameter space shown in Fig. \ref{fig:Bifurkation_reg_all} is analyzed. In this parameter region a supercritical pitchfork bifurcation of a torus is observed. The corresponding enlarged bifurcation diagram for $R=0.5$ is presented in Fig. \ref{fig:R06_bifurkation}. A solid line in $X_{1}$ direction for a fixed $\alpha$ indicates a stable torus or chaos, while individual points indicate a stable periodic invariant set.

For small values of $\alpha$ the only stable attractor in phase space is a single torus. In addition there exist two unstable periodic orbits of period $T$. An illustration of this can be seen in Fig. \ref{fig:R06_beforeafter} (upper panel) for $\alpha=0.2$. All particles end up on the torus. Note that Fig. \ref{fig:R06_beforeafter} shows the Pointcar\'{e}-section, therefore the torus appears as an invariant curve and the periodic orbits as fixed points. In this aerosol region the torus, and therefore the motion of the particles, tends to go through the local downflow regions. 

When $\alpha$ is increased, a bifurcation of the stable torus takes place at $\alpha_{crit}\approx 0.2475$. The stable torus splits into two stable tori, with an unstable torus separating the stable tori (see Fig. \ref{fig:R06_beforeafter} (lower panel) for $\alpha=0.28$). This is a supercritical pitchfork bifurcation of a torus. This change can also be seen in the Lyapunov exponents (Fig. \ref{fig:R06_bifurkation} (lower panel)), where the smaller Lyapunov exponent also becomes zero at the bifurcation point. 

Beyond the pitchfork bifurcation it depends on the initial condition to which torus the particles converge to. The unstable torus can also be seen in the basins of attraction of the stable tori, as it makes up the basin-boundary together with its stable manifolds (Fig. \ref{fig:R06_alpha0325_einz}).  

Beyond the pitchfork bifurcation the behavior gets more complicated, as parameter regions with various phase-lockings on the unstable and stable tori are encountered. Some of the larger regions with phase-locking on the stable tori can be seen in the bifurcation diagram Fig. \ref{fig:R06_bifurkation}, for example at $\alpha\approx 0.268$ and $0.305$, where the zero Lyapunov exponent becomes negative. The parameter regions with phase-locking are separated by large parameter regions of stable tori without phase-locking.

For lower values of $R$ ($R<0.4$), the behavior of the system is also much more involved and will not be examined in detail here. In this region, a number of bifurcations, in particular phase-lockings, occur on the torus before the pitchfork bifurcation. This gives rise to a complex series of bifurcations, that will be published elsewhere.

If $R$ is increased and comes closer to the bubble region ($R\rightarrow\frac{2}{3}$), another bifurcation of the stable tori can be observed. The two stable tori that were created in the pitchfork bifurcation move closer to the unstable periodic orbits and finally disappear in a Neimark-Sacker bifurcation (dashed line in Fig. \ref{fig:Bifurkation_reg_all}). There the periodic orbits become stable. All particles now move to one of these two periodic orbits, with their basins of attraction separated by the remaining unstable torus.

\subsection{\label{SubcriticalP}Subcritical Pitchfork}

\begin{figure}[htb]
		\centering
		\subfigure{\includegraphics[width=0.47\textwidth, height=5.1cm]{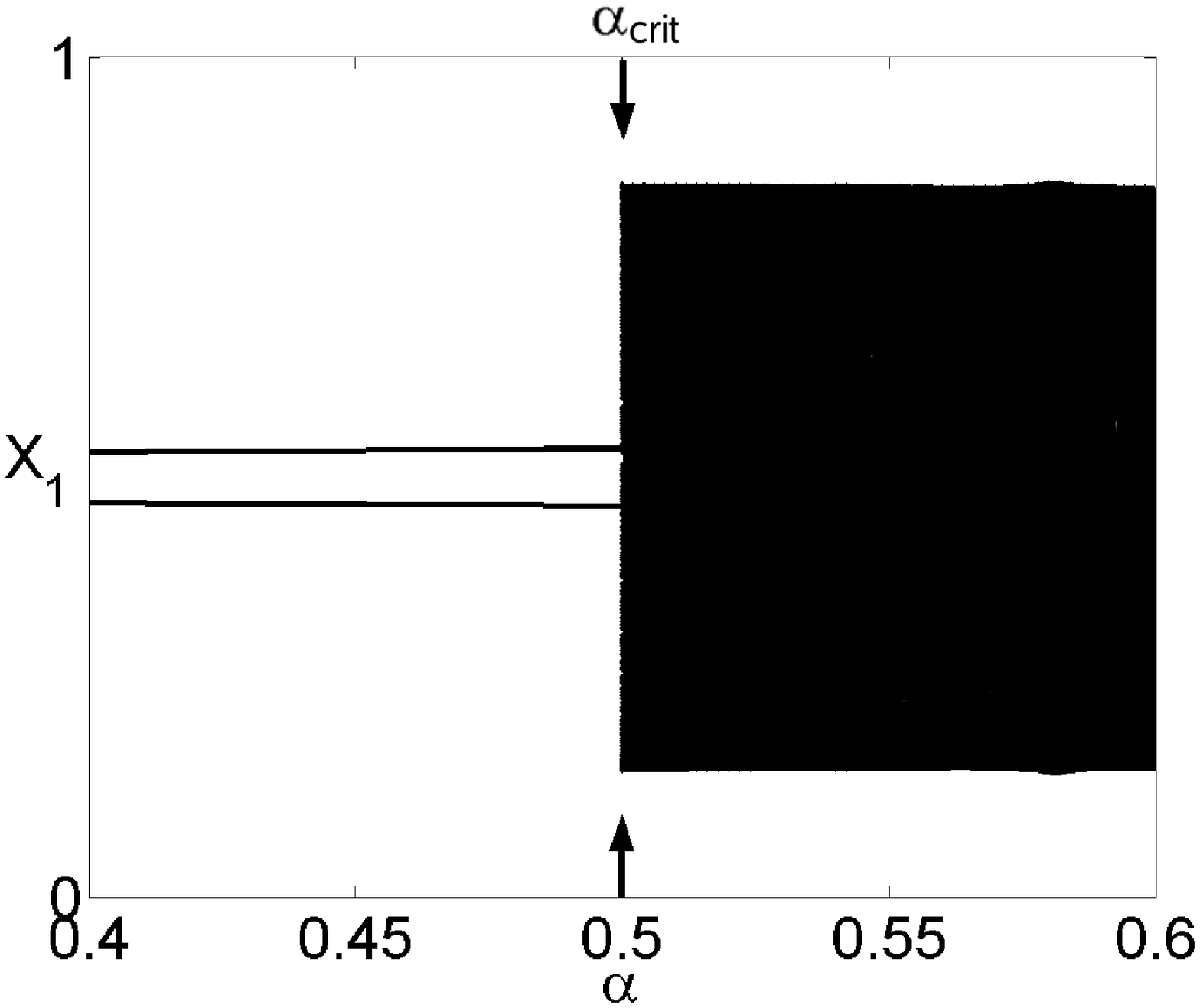}}\\
		\subfigure{\includegraphics[width=0.47\textwidth, height=5.1cm]{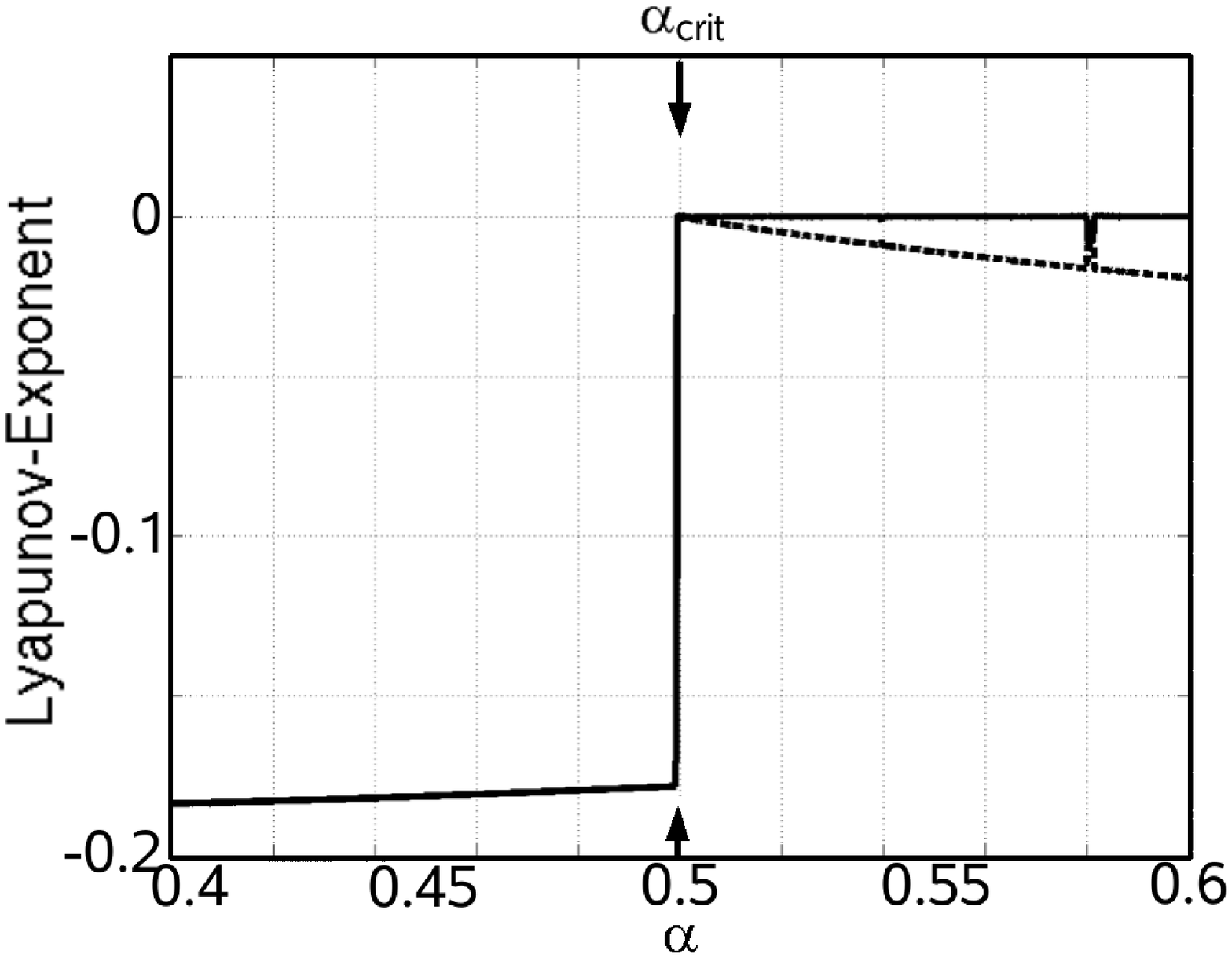}}\
		\caption{\label{fig:R1_bifurkation} Bifurcation diagram and Lyapunov exponents for $R=1$, in the area of the subcritical pitchfork bifurcation.}	
\end{figure}
\begin{figure}[htb]
		\centering
		\subfigure{\label{fig:R1_alpha04992} \includegraphics[width=0.47\textwidth, height=5.1cm]{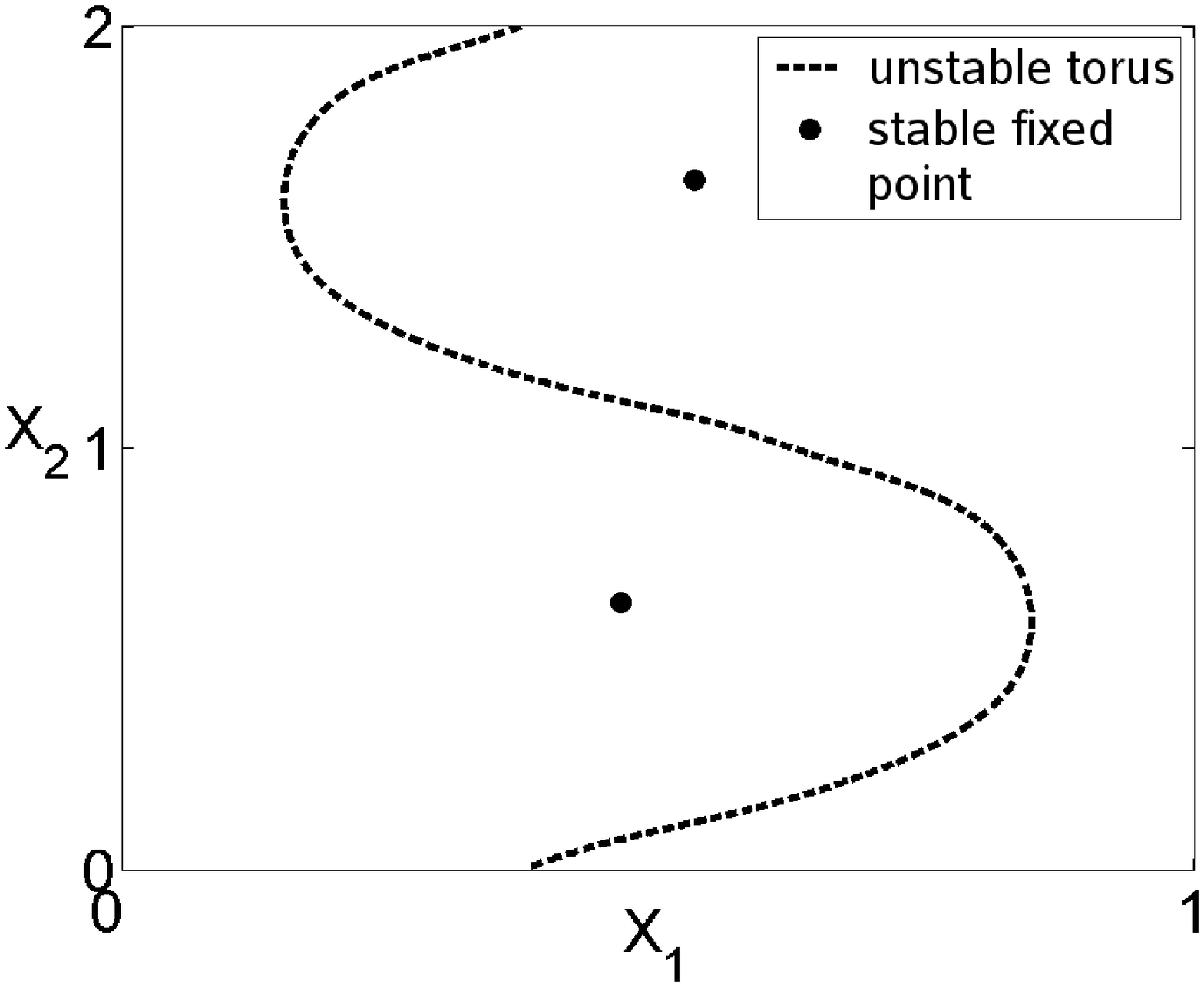}}\\
		\subfigure{\label{fig:R1_alpha051} \includegraphics[width=0.47\textwidth, height=5.1cm]{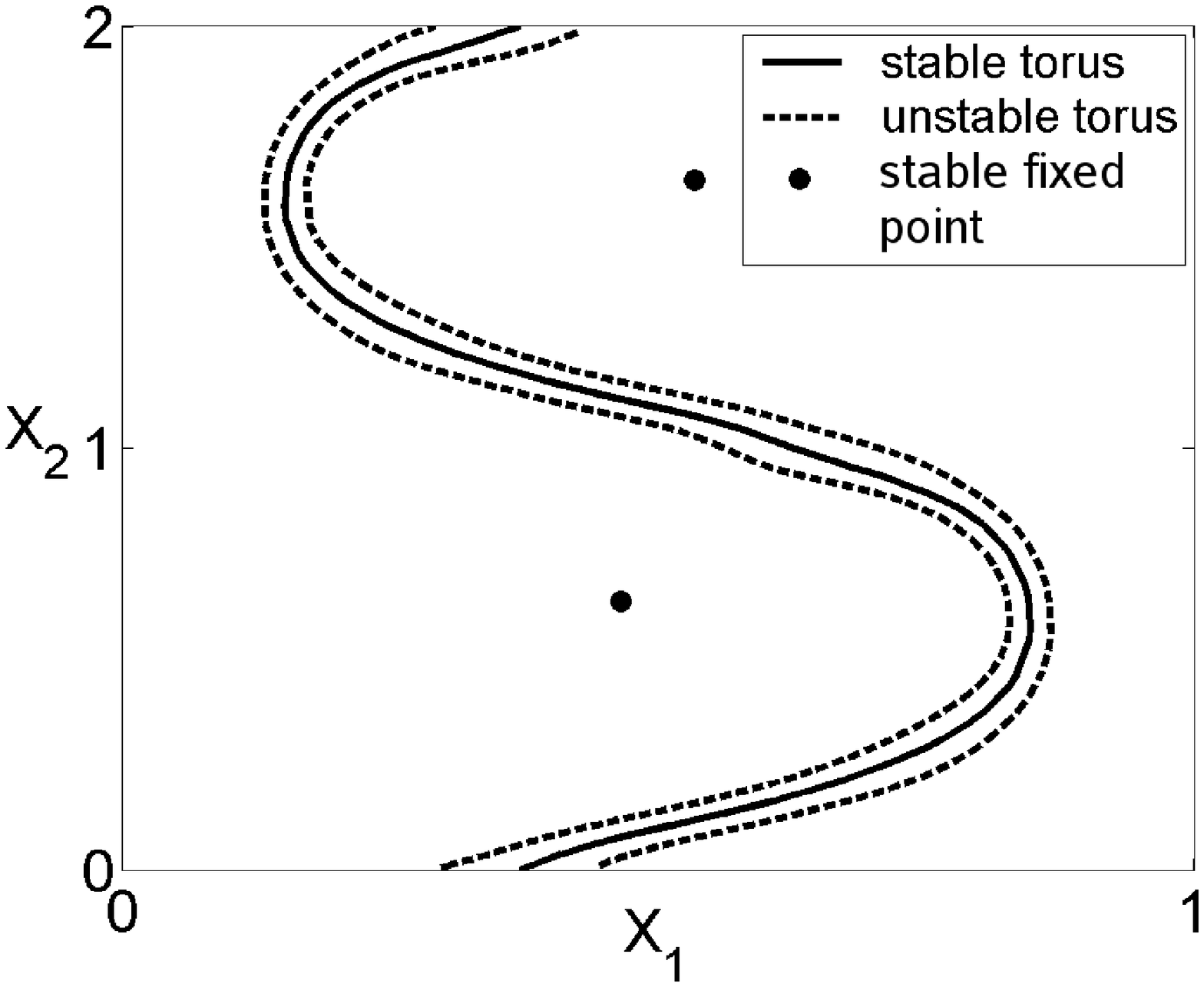}}
		\caption{\label{fig:R1_beforeafter}Subcritical pitchfork bifurcation of a torus at $R=1$ (solid line - stable torus, dashed line - unstable torus, bullet - stable periodic orbit). Upper Panel: Before the bifurcation ($\alpha=0.4992$). Lower panel: After the bifurcation ($\alpha=0.51$).}	
\end{figure}
Now we analyze the bifurcation scenarios in the other part of the parameter space, namely the bubble region with $R>\frac{2}{3}$. The dynamics is dominated by two stable fixed points, whose basins of attraction are separated by an unstable torus. However, the symmetry of the torus is reversed, with the torus now going through the regions with local upflow. 

As shown in Fig. \ref{fig:Bifurkation_reg_all} we find again a pitchfork bifurcation line for a torus, but in the bubble region this pitchfork bifurcation is subcritical. 

Let us discuss the bifurcation diagram for $R=1$ (Fig. \ref{fig:R1_bifurkation}) in the vicinity of the bifurcation. For small $\alpha$ values the long-term dynamics yields two periodic orbits of period $T$. In addition the dynamics show the existence of an unstable torus (see Fig. \ref{fig:R1_beforeafter} (upper panel)). The basins of attraction of the periodic orbits are separated by the unstable torus, resulting in a smooth basin boundary.
 
\begin{figure}[htb]
	\centering
		\includegraphics[width=0.47\textwidth, height=5.1cm]{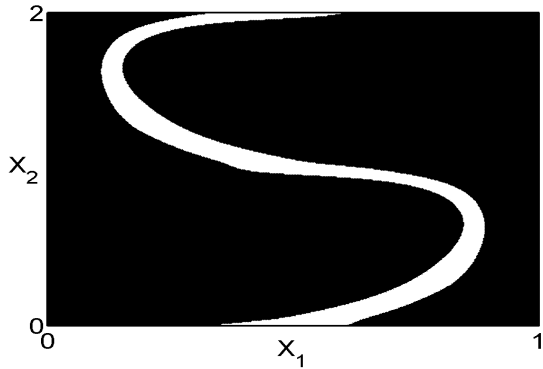}
		\caption{\label{fig:R1_alpha051_einz} Basins of attraction of the stable torus (white) and the fixed points (black) for $R=1$, $\alpha=0.51$. The basins of attraction are separated by the two unstable tori.}
 \end{figure}
At $\alpha_{crit}\approx 0.4996$ the unstable torus undergoes a bifurcation. The unstable torus becomes stable, with two new unstable tori branching off (see Fig. \ref{fig:R1_beforeafter} (lower panel) for $\alpha=0.51$). This is a subcritical pitchfork bifurcation of a torus. In the bifurcation diagram Fig. \ref{fig:R1_bifurkation} this transition is manifested by the abrupt appearance of the quasiperiodic motion indicated by the black area beyond the critical parameter value. The change can also be seen in the Lyapunov exponents, where the largest exponent $\lambda_{1}$ now is equal to zero (compare Fig. \ref{fig:Bif_Lyapunov_R1} (lower panel)).

There are now three possible states for the long-term dynamics, with the actual result for a particle depending on the initial condition. Once again, the unstable tori form the borders of the basins of attraction. In this case, the unstable tori separate the basins of attraction of the stable torus from the basins of attraction of the two periodic orbits (Fig. \ref{fig:R1_alpha051_einz}). Here, the basin boundary is again smooth.

For different values of $\alpha$, the system shows a number of other interesting types of behavior, but the analysis of these is beyond the scope of this paper and is left for further studies. Some examples include a series of phase-lockings on the unstable torus and subsequent pitchfork bifurcations of the resulting periodic orbits in the interval $\alpha\in[0.41,0.45]$, emergence of other stable periodic solutions ($\alpha>0.6$) and a break-up of the stable torus at $\alpha=1$ (see \cite{Nishikawa2002}) that gives rise to a chaotic attractor with fractal basin boundaries.

 \section{\label{Conclusion}Summary}
We have investigated the dynamics of passive finite-size particles in a fluid flow at low Reynolds numbers. It was found that the dynamics of such particles can vary greatly with their size and also with their mass relative to the fluid. For different parameter regions, periodic, quasiperiodic and chaotic motion was observed. When considering sets of particles belonging to different size classes and advected by the same flow, finite-size particles will settle on different attractors depending on their size. That means that in systems where aggregation and fragmentation processes take place the particles of different size will concentrate in different areas of the configuration space. In the case when the time span between the aggregation and fragmentation events is long compared to the time needed to approach the attractor, the particles will be distributed according to their respective attractors. But in most cases one has to expect that aggregation and fragmentation will happen more frequently, so that the transient dynamics will blur the separation of different size particles in different attractors. 

For small size classes we find a parameter range where the dynamics is dominated by quasiperiodic attractors  for particles lighter and heavier than the fluid. This quasiperiodic motion in form of a torus has been studied in more detail. The torus undergoes both a sub- and a supercritical pitchfork bifurcation, depending on the mass of the particles relative to the fluid. For particles lighter than the fluid the bifurcation is always subcritical, while for particles heavier than the fluid the bifurcation is supercritical. No continuous transition between these two bifurcations exists in this system, instead the transition between the different states happens via a Neimark-Sacker bifurcation where the two stable tori emerging in the supercritical pitchfork-bifurcation disappear. This Neimark-Sacker bifurcation forms a boundary at the line $R=\frac{2}{3}$ of neutrally buoyant particles, separating the different invariant sets observed for both kinds of pitchfork bifurcations. 

At this line of neutrally buoyant particles the symmetry of the unstable torus is reversed (compare Fig. \ref{fig:R06_beforeafter} (upper panel) and Fig. \ref{fig:R1_beforeafter} (upper panel)). This change in the symmetry can be easily understood from a physical point of view. Particles lighter than the fluid get suspended at the stable fixed points for long times, but for short times (if they start close to the unstable torus) move upwards through the flow. The torus therefore lies in the regions with local upflow. Particles heavier than the fluid sink through the fluid, the torus lies in regions with local downflow. At the line $R=\frac{2}{3}$ particles neither rise nor sink through the flow, instead they collect in the region with only horizontal fluid velocity, i.e. here the tori become a horizontal line at the border of each unit cell ($x_2=0,1,2$). 

For $R$ very close to $\frac{2}{3}$ the line of supercritical pitchfork and the line of Neimark-Sacker bifurcations seem to touch tangentially in a bifurcation of higher codimension. But the appearance of higher codimension bifurcations is beyond the scope of this paper. Additionally, for low and high values of $R$ ($R<0.4, R>1.1$) other bifurcations occur which are related to a phase-locking on the torus under consideration, giving rise to a complex network of bifurcation lines in parameter space.

As a result, a system of finite-size particles moving in a fluid flow, provides an example of a sub- and supercritical pitchfork bifurcation of a torus in a realistic physical system.
\section{Acknowledgments}
The authors wish to thank H. Osinga, A. Pikovsky, T. T\'el and M. Zaks for valuable discussions and an anonymous referee for an important remark about the chosen parameters.


\begin{thebibliography}{45}
\expandafter\ifx\csname natexlab\endcsname\relax\def\natexlab#1{#1}\fi
\expandafter\ifx\csname bibnamefont\endcsname\relax
  \def\bibnamefont#1{#1}\fi
\expandafter\ifx\csname bibfnamefont\endcsname\relax
  \def\bibfnamefont#1{#1}\fi
\expandafter\ifx\csname citenamefont\endcsname\relax
  \def\citenamefont#1{#1}\fi
\expandafter\ifx\csname url\endcsname\relax
  \def\url#1{\texttt{#1}}\fi
\expandafter\ifx\csname urlprefix\endcsname\relax\def\urlprefix{URL }\fi
\providecommand{\bibinfo}[2]{#2}
\providecommand{\eprint}[2][]{\url{#2}}

\bibitem[{\citenamefont{Aref}(1984)}]{Aref1984}
\bibinfo{author}{\bibfnamefont{H.}~\bibnamefont{Aref}},
  \bibinfo{journal}{Journal Of Fluid Mechanics} \textbf{\bibinfo{volume}{143}},
  \bibinfo{pages}{1} (\bibinfo{year}{1984}).

\bibitem[{\citenamefont{Ottino}(1989)}]{Ottino}
\bibinfo{author}{\bibfnamefont{J.}~\bibnamefont{Ottino}},
  \emph{\bibinfo{title}{The kinematics of mixing: stretching, chaos and
  transport}} (\bibinfo{publisher}{Cambridge University Press},
  \bibinfo{address}{Cambridge, U.K.}, \bibinfo{year}{1989}).

\bibitem[{\citenamefont{T\'el et~al.}(2005)\citenamefont{T\'el, DeMoura,
  Grebogi, and K\'arolyi}}]{Tel2005}
\bibinfo{author}{\bibfnamefont{T.}~\bibnamefont{T\'el}},
  \bibinfo{author}{\bibfnamefont{A.}~\bibnamefont{DeMoura}},
  \bibinfo{author}{\bibfnamefont{C.}~\bibnamefont{Grebogi}}, \bibnamefont{and}
  \bibinfo{author}{\bibnamefont{K\'arolyi}}, \bibinfo{journal}{Physics Reports}
  \textbf{\bibinfo{volume}{413}}, \bibinfo{pages}{91} (\bibinfo{year}{2005}).

\bibitem[{\citenamefont{Basset}(1888)}]{Basset}
\bibinfo{author}{\bibfnamefont{A.}~\bibnamefont{Basset}},
  \emph{\bibinfo{title}{A Treatise on Hydrodynamics}}, vol.~\bibinfo{volume}{2}
  (\bibinfo{publisher}{Deighton Bell}, \bibinfo{address}{London},
  \bibinfo{year}{1888}).

\bibitem[{\citenamefont{Boussinesq}(1903)}]{Boussinesq}
\bibinfo{author}{\bibfnamefont{J.}~\bibnamefont{Boussinesq}},
  \emph{\bibinfo{title}{Theorie Analytique de la Chaleur}},
  vol.~\bibinfo{volume}{2} (\bibinfo{publisher}{L'\'Ecole Polytechnique,
  Paris}, \bibinfo{year}{1903}).

\bibitem[{\citenamefont{Oseen}(1927)}]{Oseen}
\bibinfo{author}{\bibfnamefont{C.}~\bibnamefont{Oseen}},
  \emph{\bibinfo{title}{Hydrodynamik}} (\bibinfo{address}{Leipzig},
  \bibinfo{year}{1927}).

\bibitem[{\citenamefont{Maxey and Riley}(1983)}]{Maxey1983}
\bibinfo{author}{\bibfnamefont{M.~R.} \bibnamefont{Maxey}} \bibnamefont{and}
  \bibinfo{author}{\bibfnamefont{J.~J.} \bibnamefont{Riley}},
  \bibinfo{journal}{Phys. Fluids} \textbf{\bibinfo{volume}{26}},
  \bibinfo{pages}{883} (\bibinfo{year}{1983}).

\bibitem[{\citenamefont{Maxey}(1987)}]{Maxey1987}
\bibinfo{author}{\bibfnamefont{M.}~\bibnamefont{Maxey}},
  \bibinfo{journal}{Phys. Fluids} \textbf{\bibinfo{volume}{30}},
  \bibinfo{pages}{1915} (\bibinfo{year}{1987}).

\bibitem[{\citenamefont{Benczik et~al.}(2002)\citenamefont{Benczik, Toroczkai,
  and T\'el}}]{Benczik2002}
\bibinfo{author}{\bibfnamefont{I.~J.} \bibnamefont{Benczik}},
  \bibinfo{author}{\bibfnamefont{Z.}~\bibnamefont{Toroczkai}},
  \bibnamefont{and} \bibinfo{author}{\bibfnamefont{T.}~\bibnamefont{T\'el}},
  \bibinfo{journal}{Phys. Rev. Lett.} \textbf{\bibinfo{volume}{89}},
  \bibinfo{pages}{164501} (\bibinfo{year}{2002}).

\bibitem[{\citenamefont{Wang et~al.}(1992)\citenamefont{Wang, Maxey, Burton,
  and Stock}}]{Wang1992}
\bibinfo{author}{\bibfnamefont{L.}~\bibnamefont{Wang}},
  \bibinfo{author}{\bibfnamefont{M.}~\bibnamefont{Maxey}},
  \bibinfo{author}{\bibfnamefont{T.}~\bibnamefont{Burton}}, \bibnamefont{and}
  \bibinfo{author}{\bibfnamefont{D.}~\bibnamefont{Stock}},
  \bibinfo{journal}{Phys. Fluids A} \textbf{\bibinfo{volume}{4}},
  \bibinfo{pages}{1789} (\bibinfo{year}{1992}).

\bibitem[{\citenamefont{L\'opez}(2002)}]{Lopez2002}
\bibinfo{author}{\bibfnamefont{C.}~\bibnamefont{L\'opez}},
  \bibinfo{journal}{Phys. Rev. E} \textbf{\bibinfo{volume}{66}},
  \bibinfo{pages}{027202} (\bibinfo{year}{2002}).

\bibitem[{\citenamefont{Benczik et~al.}(2003)\citenamefont{Benczik, Toroczkai,
  and T\'el}}]{Benczik2003}
\bibinfo{author}{\bibfnamefont{I.~J.} \bibnamefont{Benczik}},
  \bibinfo{author}{\bibfnamefont{Z.}~\bibnamefont{Toroczkai}},
  \bibnamefont{and} \bibinfo{author}{\bibfnamefont{T.}~\bibnamefont{T\'el}},
  \bibinfo{journal}{Phys. Rev. E} \textbf{\bibinfo{volume}{67}},
  \bibinfo{pages}{036303} (\bibinfo{year}{2003}).

\bibitem[{\citenamefont{Do and Lai}(2004)}]{Do2004}
\bibinfo{author}{\bibfnamefont{Y.}~\bibnamefont{Do}} \bibnamefont{and}
  \bibinfo{author}{\bibfnamefont{Y.-C.} \bibnamefont{Lai}},
  \bibinfo{journal}{Phys. Rev. E} \textbf{\bibinfo{volume}{70}},
  \bibinfo{pages}{036203} (\bibinfo{year}{2004}).

\bibitem[{\citenamefont{Jackson}(1990)}]{Jackson1990}
\bibinfo{author}{\bibfnamefont{G.~A.} \bibnamefont{Jackson}},
  \bibinfo{journal}{Deep-Sea Research} \textbf{\bibinfo{volume}{37}},
  \bibinfo{pages}{1197} (\bibinfo{year}{1990}).

\bibitem[{\citenamefont{Ruiz and Izquierdo}(1997)}]{Ruiz1997}
\bibinfo{author}{\bibfnamefont{J.}~\bibnamefont{Ruiz}} \bibnamefont{and}
  \bibinfo{author}{\bibfnamefont{A.}~\bibnamefont{Izquierdo}},
  \bibinfo{journal}{Oceanologica Acta} \textbf{\bibinfo{volume}{20}},
  \bibinfo{pages}{597} (\bibinfo{year}{1997}).

\bibitem[{\citenamefont{Melis et~al.}(1999)\citenamefont{Melis, Verduyn,
  Storti, Morbidelli, and Baldyga}}]{Melis1999}
\bibinfo{author}{\bibfnamefont{S.}~\bibnamefont{Melis}},
  \bibinfo{author}{\bibfnamefont{M.}~\bibnamefont{Verduyn}},
  \bibinfo{author}{\bibfnamefont{G.}~\bibnamefont{Storti}},
  \bibinfo{author}{\bibfnamefont{M.}~\bibnamefont{Morbidelli}},
  \bibnamefont{and} \bibinfo{author}{\bibfnamefont{J.}~\bibnamefont{Baldyga}},
  \bibinfo{journal}{AIChE Journal} \textbf{\bibinfo{volume}{45}},
  \bibinfo{pages}{1383} (\bibinfo{year}{1999}).

\bibitem[{\citenamefont{Thomas et~al.}(1999)\citenamefont{Thomas, Judd, and
  Fawcett}}]{Thomas1999}
\bibinfo{author}{\bibfnamefont{D.}~\bibnamefont{Thomas}},
  \bibinfo{author}{\bibfnamefont{S.}~\bibnamefont{Judd}}, \bibnamefont{and}
  \bibinfo{author}{\bibfnamefont{N.}~\bibnamefont{Fawcett}},
  \bibinfo{journal}{Water Research} \textbf{\bibinfo{volume}{33}},
  \bibinfo{pages}{1579} (\bibinfo{year}{1999}).

\bibitem[{\citenamefont{Flesch et~al.}(1999)\citenamefont{Flesch, Spicer, and
  Pratsinis}}]{Flesch1999}
\bibinfo{author}{\bibfnamefont{J.}~\bibnamefont{Flesch}},
  \bibinfo{author}{\bibfnamefont{P.}~\bibnamefont{Spicer}}, \bibnamefont{and}
  \bibinfo{author}{\bibfnamefont{S.}~\bibnamefont{Pratsinis}},
  \bibinfo{journal}{AIChE Journal} \textbf{\bibinfo{volume}{45}},
  \bibinfo{pages}{1114} (\bibinfo{year}{1999}).

\bibitem[{\citenamefont{Han et~al.}(2003)\citenamefont{Han, Akeprathumchai,
  Wickramasinghe, and Qian}}]{Binbing2003}
\bibinfo{author}{\bibfnamefont{B.}~\bibnamefont{Han}},
  \bibinfo{author}{\bibfnamefont{S.}~\bibnamefont{Akeprathumchai}},
  \bibinfo{author}{\bibfnamefont{S.}~\bibnamefont{Wickramasinghe}},
  \bibnamefont{and} \bibinfo{author}{\bibfnamefont{X.}~\bibnamefont{Qian}},
  \bibinfo{journal}{AIChE Journal} \textbf{\bibinfo{volume}{49}},
  \bibinfo{pages}{1687} (\bibinfo{year}{2003}).

\bibitem[{\citenamefont{Nishikawa et~al.}(2001)\citenamefont{Nishikawa,
  Toroczkai, and Grebogi}}]{Nishikawa2001}
\bibinfo{author}{\bibfnamefont{T.}~\bibnamefont{Nishikawa}},
  \bibinfo{author}{\bibfnamefont{Z.}~\bibnamefont{Toroczkai}},
  \bibnamefont{and} \bibinfo{author}{\bibfnamefont{C.}~\bibnamefont{Grebogi}},
  \bibinfo{journal}{Phys. Rev. Lett.} \textbf{\bibinfo{volume}{87}},
  \bibinfo{pages}{038301} (\bibinfo{year}{2001}).

\bibitem[{\citenamefont{Nishikawa et~al.}(2002)\citenamefont{Nishikawa,
  Toroczkai, Grebogi, and T\'el}}]{Nishikawa2002}
\bibinfo{author}{\bibfnamefont{T.}~\bibnamefont{Nishikawa}},
  \bibinfo{author}{\bibfnamefont{Z.}~\bibnamefont{Toroczkai}},
  \bibinfo{author}{\bibfnamefont{C.}~\bibnamefont{Grebogi}}, \bibnamefont{and}
  \bibinfo{author}{\bibfnamefont{T.}~\bibnamefont{T\'el}},
  \bibinfo{journal}{Phys. Rev. E} \textbf{\bibinfo{volume}{65}},
  \bibinfo{pages}{026216} (\bibinfo{year}{2002}).

\bibitem[{\citenamefont{Yu et~al.}(1990)\citenamefont{Yu, Grebogi, and
  Ott}}]{Yu1990}
\bibinfo{author}{\bibfnamefont{L.}~\bibnamefont{Yu}},
  \bibinfo{author}{\bibfnamefont{C.}~\bibnamefont{Grebogi}}, \bibnamefont{and}
  \bibinfo{author}{\bibfnamefont{E.}~\bibnamefont{Ott}}, in
  \emph{\bibinfo{booktitle}{Nonlinear Structure in Physical Systems}}, edited
  by \bibinfo{editor}{\bibfnamefont{L.}~\bibnamefont{Lam}} \bibnamefont{and}
  \bibinfo{editor}{\bibfnamefont{H.}~\bibnamefont{Morris}}
  (\bibinfo{publisher}{Springer Verlag}, \bibinfo{address}{New York},
  \bibinfo{year}{1990}), pp. \bibinfo{pages}{223--231}.

\bibitem[{\citenamefont{Broer et~al.}(1994)\citenamefont{Broer, Huitema,
  Takens, and Braaksma}}]{UnfoldingsBifurcationsofTori}
\bibinfo{author}{\bibfnamefont{H.}~\bibnamefont{Broer}},
  \bibinfo{author}{\bibfnamefont{G.}~\bibnamefont{Huitema}},
  \bibinfo{author}{\bibfnamefont{F.}~\bibnamefont{Takens}}, \bibnamefont{and}
  \bibinfo{author}{\bibfnamefont{B.}~\bibnamefont{Braaksma}},
  \emph{\bibinfo{title}{Unfoldings and bifurcations of quasi-periodic tori}},
  vol.~\bibinfo{volume}{83} of \emph{\bibinfo{series}{Memoirs of the American
  Methematical Society}} (\bibinfo{publisher}{American Mathematical Society},
  \bibinfo{address}{Providence, Rhode Island, USA}, \bibinfo{year}{1994}).

\bibitem[{\citenamefont{Newhouse et~al.}(1978)\citenamefont{Newhouse, Ruelle,
  and Takens}}]{Newhouse1978}
\bibinfo{author}{\bibfnamefont{S.}~\bibnamefont{Newhouse}},
  \bibinfo{author}{\bibfnamefont{D.}~\bibnamefont{Ruelle}}, \bibnamefont{and}
  \bibinfo{author}{\bibfnamefont{F.}~\bibnamefont{Takens}},
  \bibinfo{journal}{Commun. Math. Phys.} \textbf{\bibinfo{volume}{64}},
  \bibinfo{pages}{35} (\bibinfo{year}{1978}).

\bibitem[{\citenamefont{Afraimovich and Shilnikov}(1983)}]{Afraimovich1983}
\bibinfo{author}{\bibfnamefont{V.}~\bibnamefont{Afraimovich}} \bibnamefont{and}
  \bibinfo{author}{\bibfnamefont{L.}~\bibnamefont{Shilnikov}}, in
  \emph{\bibinfo{booktitle}{Methods of qualitative theory of differential
  equations}} (\bibinfo{publisher}{Gorkii University}, \bibinfo{address}{Gorky,
  Russia}, \bibinfo{year}{1983}), pp. \bibinfo{pages}{3--26}.

\bibitem[{\citenamefont{Ostlund et~al.}(1983)\citenamefont{Ostlund, Rand,
  Sethna, and Siggia}}]{Ostlund1983}
\bibinfo{author}{\bibfnamefont{S.}~\bibnamefont{Ostlund}},
  \bibinfo{author}{\bibfnamefont{D.}~\bibnamefont{Rand}},
  \bibinfo{author}{\bibfnamefont{J.}~\bibnamefont{Sethna}}, \bibnamefont{and}
  \bibinfo{author}{\bibfnamefont{E.}~\bibnamefont{Siggia}},
  \bibinfo{journal}{Physica D} \textbf{\bibinfo{volume}{8}},
  \bibinfo{pages}{303} (\bibinfo{year}{1983}).

\bibitem[{\citenamefont{Anishchenko and Safonova}(1987)}]{Anishchenko1987}
\bibinfo{author}{\bibfnamefont{V.~S.} \bibnamefont{Anishchenko}}
  \bibnamefont{and} \bibinfo{author}{\bibfnamefont{M.~A.}
  \bibnamefont{Safonova}}, \bibinfo{journal}{Radiotekhnika I Elektronika}
  \textbf{\bibinfo{volume}{32}}, \bibinfo{pages}{1207} (\bibinfo{year}{1987}).

\bibitem[{\citenamefont{Baesens et~al.}(1991)\citenamefont{Baesens,
  Guckenheimer, Kim, and MacKay}}]{Baesens1991}
\bibinfo{author}{\bibfnamefont{C.}~\bibnamefont{Baesens}},
  \bibinfo{author}{\bibfnamefont{J.}~\bibnamefont{Guckenheimer}},
  \bibinfo{author}{\bibfnamefont{S.}~\bibnamefont{Kim}}, \bibnamefont{and}
  \bibinfo{author}{\bibfnamefont{R.}~\bibnamefont{MacKay}},
  \bibinfo{journal}{Physica D} \textbf{\bibinfo{volume}{49}},
  \bibinfo{pages}{387} (\bibinfo{year}{1991}).

\bibitem[{\citenamefont{Held and Jeffries}(1986)}]{Held1986}
\bibinfo{author}{\bibfnamefont{G.}~\bibnamefont{Held}} \bibnamefont{and}
  \bibinfo{author}{\bibfnamefont{C.}~\bibnamefont{Jeffries}},
  \bibinfo{journal}{Physical Review Letters} \textbf{\bibinfo{volume}{56}},
  \bibinfo{pages}{1183} (\bibinfo{year}{1986}).

\bibitem[{\citenamefont{Schneider et~al.}(1993)\citenamefont{Schneider, Heng,
  and Martienssen}}]{Schneider1993}
\bibinfo{author}{\bibfnamefont{T.}~\bibnamefont{Schneider}},
  \bibinfo{author}{\bibfnamefont{H.}~\bibnamefont{Heng}}, \bibnamefont{and}
  \bibinfo{author}{\bibfnamefont{W.}~\bibnamefont{Martienssen}},
  \bibinfo{journal}{Europhysics Letters} \textbf{\bibinfo{volume}{22}},
  \bibinfo{pages}{499} (\bibinfo{year}{1993}).

\bibitem[{\citenamefont{Giberti and Zansai}(1992)}]{Giberti1993}
\bibinfo{author}{\bibfnamefont{C.}~\bibnamefont{Giberti}} \bibnamefont{and}
  \bibinfo{author}{\bibfnamefont{R.}~\bibnamefont{Zansai}},
  \bibinfo{journal}{Physica D} \textbf{\bibinfo{volume}{65}},
  \bibinfo{pages}{300} (\bibinfo{year}{1992}).

\bibitem[{\citenamefont{Anishchenko et~al.}(1994)\citenamefont{Anishchenko,
  Safonova, Feudel, and Kurths}}]{Anishchenko1993}
\bibinfo{author}{\bibfnamefont{V.}~\bibnamefont{Anishchenko}},
  \bibinfo{author}{\bibfnamefont{M.}~\bibnamefont{Safonova}},
  \bibinfo{author}{\bibfnamefont{U.}~\bibnamefont{Feudel}}, \bibnamefont{and}
  \bibinfo{author}{\bibfnamefont{J.}~\bibnamefont{Kurths}},
  \bibinfo{journal}{Int. J. Bifurcation and Chaos}
  \textbf{\bibinfo{volume}{4}}, \bibinfo{pages}{595} (\bibinfo{year}{1994}).

\bibitem[{\citenamefont{Feudel et~al.}(1996)\citenamefont{Feudel, Safonova,
  Kurths, and Anishchenko}}]{Feudel1996}
\bibinfo{author}{\bibfnamefont{U.}~\bibnamefont{Feudel}},
  \bibinfo{author}{\bibfnamefont{M.}~\bibnamefont{Safonova}},
  \bibinfo{author}{\bibfnamefont{J.}~\bibnamefont{Kurths}}, \bibnamefont{and}
  \bibinfo{author}{\bibfnamefont{V.}~\bibnamefont{Anishchenko}},
  \bibinfo{journal}{Int. J. Bifurcation and Chaos}
  \textbf{\bibinfo{volume}{6}}, \bibinfo{pages}{1319} (\bibinfo{year}{1996}).

\bibitem[{\citenamefont{Francheschini and Zanasi}(1992)}]{Franceschini1992}
\bibinfo{author}{\bibfnamefont{V.}~\bibnamefont{Francheschini}}
  \bibnamefont{and} \bibinfo{author}{\bibfnamefont{R.}~\bibnamefont{Zanasi}},
  \bibinfo{journal}{Nonlinearity} \textbf{\bibinfo{volume}{4}},
  \bibinfo{pages}{189} (\bibinfo{year}{1992}).

\bibitem[{\citenamefont{Belogortsev et~al.}(1993)\citenamefont{Belogortsev,
  Vavriv, and Tretyakov}}]{Belogortsev1993}
\bibinfo{author}{\bibfnamefont{A.~B.} \bibnamefont{Belogortsev}},
  \bibinfo{author}{\bibfnamefont{D.~M.} \bibnamefont{Vavriv}},
  \bibnamefont{and} \bibinfo{author}{\bibfnamefont{O.~A.}
  \bibnamefont{Tretyakov}}, \bibinfo{journal}{Applied Mechanics Reviews}
  \textbf{\bibinfo{volume}{46}}, \bibinfo{pages}{374} (\bibinfo{year}{1993}).

\bibitem[{\citenamefont{Feudel and Seehafer}(1995)}]{Feudel1995}
\bibinfo{author}{\bibfnamefont{F.}~\bibnamefont{Feudel}} \bibnamefont{and}
  \bibinfo{author}{\bibfnamefont{N.}~\bibnamefont{Seehafer}},
  \bibinfo{journal}{Phys. Rev. E} \textbf{\bibinfo{volume}{52}},
  \bibinfo{pages}{3506} (\bibinfo{year}{1995}).

\bibitem[{\citenamefont{Sun}(2003)}]{Sun2003}
\bibinfo{author}{\bibfnamefont{J.~H.} \bibnamefont{Sun}},
  \bibinfo{journal}{Acta Mathematica Sinicia, English Series}
  \textbf{\bibinfo{volume}{19}}, \bibinfo{pages}{159} (\bibinfo{year}{2003}).

\bibitem[{\citenamefont{Chandrasekhar}(1961)}]{Chandrasekhar}
\bibinfo{author}{\bibfnamefont{S.}~\bibnamefont{Chandrasekhar}},
  \emph{\bibinfo{title}{Hydrodynamic and Hydromagnetic Stability}}
  (\bibinfo{publisher}{Oxford University Press, Elt House, London W. 1},
  \bibinfo{year}{1961}).

\bibitem[{\citenamefont{Wiggins}(1988)}]{WigginsGlobalChaos}
\bibinfo{author}{\bibfnamefont{S.}~\bibnamefont{Wiggins}},
  \emph{\bibinfo{title}{Global bifurcations and chaos: analytical methods}}
  (\bibinfo{publisher}{Springer Verlag}, \bibinfo{address}{New York},
  \bibinfo{year}{1988}).

\bibitem[{\citenamefont{Feudel and Jansen}(1992)}]{Feudel1992}
\bibinfo{author}{\bibfnamefont{U.}~\bibnamefont{Feudel}} \bibnamefont{and}
  \bibinfo{author}{\bibfnamefont{W.}~\bibnamefont{Jansen}},
  \bibinfo{journal}{International Journal of Bifurcation and Chaos}
  \textbf{\bibinfo{volume}{2}}, \bibinfo{pages}{773} (\bibinfo{year}{1992}).

\bibitem[{\citenamefont{Hairer et~al.}(1993)\citenamefont{Hairer, Norsett, and
  Wanner}}]{ODE1}
\bibinfo{author}{\bibfnamefont{E.}~\bibnamefont{Hairer}},
  \bibinfo{author}{\bibfnamefont{S.}~\bibnamefont{Norsett}}, \bibnamefont{and}
  \bibinfo{author}{\bibfnamefont{G.}~\bibnamefont{Wanner}},
  \emph{\bibinfo{title}{Solving Ordinary Differential Equations I: Nonstiff
  Problems}}, vol.~\bibinfo{volume}{8} of \emph{\bibinfo{series}{Springer
  Series in Comput. Math.}} (\bibinfo{publisher}{Springer Verlag},
  \bibinfo{address}{New York}, \bibinfo{year}{1993}), \bibinfo{edition}{2nd}
  ed.

\bibitem[{\citenamefont{Manton}(1974)}]{Manton1974}
\bibinfo{author}{\bibfnamefont{M.}~\bibnamefont{Manton}},
  \bibinfo{journal}{Boundary-Layer Meteorol.} \textbf{\bibinfo{volume}{6}},
  \bibinfo{pages}{487} (\bibinfo{year}{1974}).

\bibitem[{\citenamefont{Nusse and Yorke}(1989)}]{Nusse-Yorke-89}
\bibinfo{author}{\bibfnamefont{H.}~\bibnamefont{Nusse}} \bibnamefont{and}
  \bibinfo{author}{\bibfnamefont{J.}~\bibnamefont{Yorke}},
  \bibinfo{journal}{Physica D} \textbf{\bibinfo{volume}{36}},
  \bibinfo{pages}{137} (\bibinfo{year}{1989}).

\bibitem[{\citenamefont{Shimada and Nagashima}(1979)}]{Shimada-Nagashima-79}
\bibinfo{author}{\bibfnamefont{I.}~\bibnamefont{Shimada}} \bibnamefont{and}
  \bibinfo{author}{\bibfnamefont{T.}~\bibnamefont{Nagashima}},
  \bibinfo{journal}{Progr. Theor. Phys.} \textbf{\bibinfo{volume}{61}},
  \bibinfo{pages}{1605} (\bibinfo{year}{1979}).

\bibitem[{\citenamefont{Osinga et~al.}(2001)\citenamefont{Osinga, Wiersing,
  Glendinning, and Feudel}}]{Osinga2001}
\bibinfo{author}{\bibfnamefont{H.}~\bibnamefont{Osinga}},
  \bibinfo{author}{\bibfnamefont{J.}~\bibnamefont{Wiersing}},
  \bibinfo{author}{\bibfnamefont{P.}~\bibnamefont{Glendinning}},
  \bibnamefont{and} \bibinfo{author}{\bibfnamefont{U.}~\bibnamefont{Feudel}},
  \bibinfo{journal}{Int. J. Bifurcation and Chaos}
  \textbf{\bibinfo{volume}{11}}, \bibinfo{pages}{3085} (\bibinfo{year}{2001}).

\end{thebibliography}
\end{document}